\documentclass[aps,prb,preprint,superscriptaddress,showpacs,floatfix]{revtex4}
\usepackage{hyperref}
\usepackage{amsmath,amssymb,amsfonts}
\usepackage{graphicx}
\usepackage{color}
\DeclareGraphicsExtensions{.jpg,.pdf,.png,.eps}

\begin{document}
\newcommand{\CaP} {CaFe$_{2}$As$_{2}$ }
\newcommand{\CaCo} {Ca(Fe$_{1-x}$Co$_{x}$)$_{2}$As$_{2}$ }
\newcommand{\CaNi} {Ca(Fe$_{1-x}$Ni$_{x}$)$_{2}$As$_{2}$ }
\newcommand{\CaRh} {Ca(Fe$_{1-x}$Rh$_{x}$)$_{2}$As$_{2}$ }

\title{Combined effects of transition metal (Ni and Rh) substitution and annealing/quenching on physical properties of CaFe$_{2}$As$_{2}$}

\author{S.~Ran}
\affiliation{Ames Laboratory, Iowa State University, Ames, Iowa 50011, USA}
\affiliation{Department of Physics and Astronomy, Iowa State University, Ames, Iowa 50011, USA}
\author{S.~L.~Bud'ko}
\affiliation{Ames Laboratory, Iowa State University, Ames, Iowa 50011, USA}
\affiliation{Department of Physics and Astronomy, Iowa State University, Ames, Iowa 50011, USA}
\author{W.~E.~Straszheim}
\affiliation{Ames Laboratory, Iowa State University, Ames, Iowa 50011, USA}
\author{P.~C.~Canfield}
\affiliation{Ames Laboratory, Iowa State University, Ames, Iowa 50011, USA}
\affiliation{Department of Physics and Astronomy, Iowa State University, Ames, Iowa 50011, USA}
\date{\today}

\begin{abstract}

We performed systematic studies of the combined effects of annealing/quenching temperature ({\itshape T}$_{A/Q}$) and T = Ni, Rh substitution ({\itshape x}) on the physical properties of Ca(Fe$_{1-x}$T$_{x}$)$_{2}$As$_{2}$. We constructed two-dimensional, {\itshape T}$_{A/Q}$-{\itshape x} phase diagrams for the low-temperature states for both substitutions to map out the relations between ground states and compared them with that of Co-substitution. Ni-substitution, which brings one more extra electron per substituted atom and suppresses the {\itshape c}-lattice parameter at roughly the same rate as Co-substitution, leads to a similar parameter range of antiferromagnetic/orthorhombic in the {\itshape T}$_{A/Q}$-{\itshape x} space as that found for Co-substitution, but has the parameter range for superconductivity shrunk (roughly by a factor of two). This result is similar to what is found when Co- and Ni-substituted BaFe$_{2}$As$_{2}$ are compared. On the other hand, Rh-substitution, which brings the same amount of extra electrons as does Co-substitution, but suppresses the {\itshape c}-lattice parameter more rapidly, has a different phase diagram. The collapsed tetragonal phase exists much more pervasively, to the exclusion of the normal, paramagnetic, tetragonal phase. The range of antiferromagnetic/orthorhombic phase space is noticeably reduced, and the superconducting region is substantially suppressed, essentially truncated by the collapsed tetragonal phase. In addition, we found that whereas for Co-substitution there was no difference between phase diagrams for samples annealed for one or seven days, for Ni- and Rh- substitutions a second, reversible, effect of annealing was revealed by seven-day anneals.

\end{abstract}

\maketitle

\section{Introduction}

Among the parent compounds of the Fe-based superconductors, \CaP manifests unique physical properties and has become a model system for understanding high-{\itshape T}$_{c}$ superconductivity in Fe-based superconductors.\cite{Ni08Ca,Goldman08,Kreyssig08,Goldman09,Torikachvili08Ca,Yu09,Canfield09Ca,Lee08,Prokes10,Park08,Ran11,Ran12} The magnetic and structural phase transitions are strongly coupled and first order with hysteresis of several degrees as seen in thermodynamic, transport, and microscopic measurements.\cite{Ni08Ca,Goldman08} Also, \CaP is the most pressure sensitive of the AFe$_{2}$As$_{2}$ (A = Ba, Sr, Ca) and 1111 compounds with its antiferromagnetic/orthorhombic (AFM/ORTH) phase transition being initially suppressed by over 100~K per GPa and a then non-moment bearing, collapsed tetragonal (cT) phase being stabilized by $\sim$0.4~GPa.\cite{Kreyssig08,Goldman09,Torikachvili08Ca,Yu09,Canfield09Ca,Lee08,Prokes10,Park08}

Previous work shows that the phase transition temperatures and even ground state of \CaP can be controlled and tuned by post-growth annealing and quenching of single crystal samples.\cite{Ran11} Crystals of \CaP grown from Sn-flux and quenched from 600$^\circ$C manifest AFM/ORTH phase transition at 170 K. On the other hand, we found that crystals grown from FeAs-rich solutions, quenched from higher temperature (960$^\circ$C) exhibit a transition to the cT phase below 100~K at ambient pressure. Further, we found that, for the FeAs-flux grown samples, a process of post-growth annealing and quenching can be used as an additional control parameter to tune the ground state of \CaP systematically, suppressing the AFM/ORTH transition temperature from 170 K to $\sim$ 100 K and then stabilizing an ambient pressure cT phase by changing {\itshape T}$_{A/Q}$ from 350$^\circ$C to 800 and further to 960$^\circ$C. Time dependence of annealing effects was studied to establish annealing protocols and the {\itshape T}-{\itshape T}$_{A/Q}$ phase diagram was constructed. It was found that the {\itshape T}-{\itshape T}$_{A/Q}$ and the {\itshape T}-{\itshape P} phase diagrams are similar, suggesting that the effect of annealing and quenching is similar to that of pressure.\cite{Ran11,Dhaka14} Based on the TEM results, which reveal nano-scale, low strain precipitates in the sample with {\itshape T}$_{A/Q}$ = 500$^\circ$C and a strain field in the sample with {\itshape T}$_{A/Q}$ = 960$^\circ$C, it is likely that the annealing and quenching process controls how a small excess of FeAs is brought in and out of the CaFe$_{2}$As$_{2}$ matrix and, as a result, the amount of stress and strain built up in the samples, mimicing the modest pressures needed to stabilize the cT phase.

Having mastered the control of FeAs grown CaFe$_{2}$As$_{2}$, we are able to tune the ground state of this system with the combination of chemical substitution, annealing/quenching and application of hydrostatic pressure.\cite{Ran12,Sergey13,Gati12} The combined effects of Co-substitution and annealing/quenching on the physical properties of \CaP were studied and a 3D phase diagram, with Co concentration, {\itshape x}, and annealing/quenching temperature, {\itshape T}$_{A/Q}$, as two independent control parameters, was constructed. At ambient pressure, the \CaCo system offers ready access to the salient low-temperature states associated with Fe-based superconductors: antiferromagnetic/orthorhombic (AFM/ORTH), superconducting/paramagnetic/tetragonal (SC/PM/T), non superconducting/paramagnetic/tetragonal (N/PM/T) and non-moment bearing/collapsed tetragonal (cT). In a similar manner, for {\itshape x} = 0.028, very modest, hydrostatic pressure ({\itshape P} $<$ 260 MPa) can change the low-temperature ground state from AFM/ORTH to SC/PM/T, to a cT phase. Detailed comparisons of the {\itshape P} and {\itshape T}$_{A/Q}$ dependence of the \CaCo phase diagrams further supported the similarity of the effects of {\itshape T}$_{A/Q}$ and {\itshape P} on these compounds. For example, for {\itshape x} = 0.028, a scaling of $\Delta${\itshape T}$_{A/Q}$ = 100$^\circ$C being equivalent to $\Delta${\itshape P} = 85 MPa allowed for the {\itshape T}-{\itshape T}$_{A/Q}$ and the {\itshape T}-{\itshape P} phase diagrams to fall on a single manifold.\cite{Gati12}

In the case of BaFe$_{2}$As$_{2}$ system,\cite{Ni09TM,Canfield09TM,Ni10TM,Canfield10} comparison of the phase diagrams of various transition metal substitutions reveals that while the suppression of the structural/magnetic phase transition scales, roughly, with impurity concentration, {\itshape x}, the superconductivity is rather controlled by extra electron count, {\itshape e}. Steric effect seems not to play any important role in determining the phase diagram, with Co- (Ni-) and Rh- (Pd-) substitution having exceptionally similar effect, especially on the superconducting dome on the overdoped side.

In order to compare the phase diagrams of various transition metal substitutions in CaFe$_{2}$As$_{2}$, we expand the exploration of transition metal substitution to Ni and Rh. Compared with Co-substitution, Ni-substitution brings one more extra electron per substituted atom, while suppressing {\itshape c}-lattice parameter in a very similar manner. On the other hand, Rh-substitution brings nominally the same amount of extra electrons as Co-substitution, although from a 4d-shell rather than a 3d-shell, while suppressing the {\itshape c}-lattice parameter much more rapidly. Therefore, comparing Co-substitution with Ni- and Rh-substitution will potentially help us understand the changes of physical properties of \CaP system caused by (i) band filling and (ii) steric effect. As we will show, this is more complicated than in the case of Ba(Fe$_{1-x}$T$_{x}$)$_{2}$As$_{2}$, not only due to the existence of one more control parameter, {\itshape T}$_{A/Q}$, but also because \CaP is much more sensitive to the pressure, and therefore to the steric effect. 

Due to the fact that two independent control parameters ({\itshape T}$_{A/Q}$ and {\itshape x}) define this phase space, large amounts of temperature dependent data were collected and used to assemble the various phase diagrams. In the main body of this paper only selected sets of data will be presented and the rest of data will be presented in the Appendix. 

\section{Experimental methods}

Single crystals of Ca(Fe$_{1-x}$TM$_{x}$)$_{2}$As$_{2}$ were grown out of FeAs-flux, using conventional high-temperature solution growth techniques.\cite{Ran11,Fisk89,Canfield92,Canfieldbook,Sefat08} The growth protocol and the post-growth thermal treatments for both Ni- and Rh-substituted CaFe$_{2}$As$_{2}$ are the same as for Co-substitution.\cite{Ran12} In the process of decanting off the excess flux, the samples were essentially quenched from 960$^\circ$C to room temperature, which, according to our previous study, causes strain inside the samples, leading to behavior different from Sn grown samples.\cite{Ni08Ca,Hu11Ca} These samples will be referred to as {\itshape T}$_{A/Q}$ = 960$^\circ$C (as grown) samples. post-growth, thermal treatments of samples involve annealing samples at a certain temperature ranging from 350$^\circ$C to 800$^\circ$C and subsequently quenching from this annealing temperature to room temperature. These samples will be identified as {\itshape T}$_{A/Q}$ = 350$^\circ$C to {\itshape T}$_{A/Q}$ = 800$^\circ$C. The data presented in the first part of the text were all collected on the samples annealed for 24 hours. As will be discussed in the second part of this paper, for Ni- and Rh-substitution, longer annealing times appear to further introduce a second, controllable process. Details about the annealing and quenching technique can be found in references.\cite{Ran11,Ran12}.

Elemental analysis was performed on each of these batches using wavelength-dispersive x-ray spectroscopy (WDS) in the electron probe microanalyzer of a JEOL JXA-8200 electron microprobe. Since the properties of a given sample are found to be determined by both the transition metal substitution level, {\itshape x}, and the post-growth annealing/quenching temperature, {\itshape T}$_{A/Q}$, samples are fully identified by providing both of these parameters. 

Diffraction measurements on the platelike samples were performed at room temperature using a Rigaku Miniflex diffractometer with Cu {\itshape K}$\alpha$ radiation. Only (00l) peaks are observed from which the values of the {\itshape c}-lattice parameter are inferred. Standard powder x-ray diffraction was not attempted because grinding of \CaP leads to severe peak broadening and even changes in physical properties, as has been outlined in our previous work. \cite{Ni08Ca,Ran11,Ran12} 

Temperature dependent DC magnetization measurements were made in Quantum Design Magnetic Property Measurement Systems (MPMS). The in-plane, temperature dependent, electrical resistance measurements were performed in Quantum Design Physical Property Measurement Systems (PPMS) or in Quantum Design MPMS systems operated in external device control (EDC) mode, in conjunction with Linear Research LR700 four-probe AC resistance bridges (f =16Hz, I = 1mA). The electrical contacts were placed on the samples in standard 4-probe geometry, using Pt wires attached to a sample surface with Epotek H20E silver epoxy.

\begin{figure}[!htbp]
\begin{center}
\includegraphics[angle=0,width=100mm]{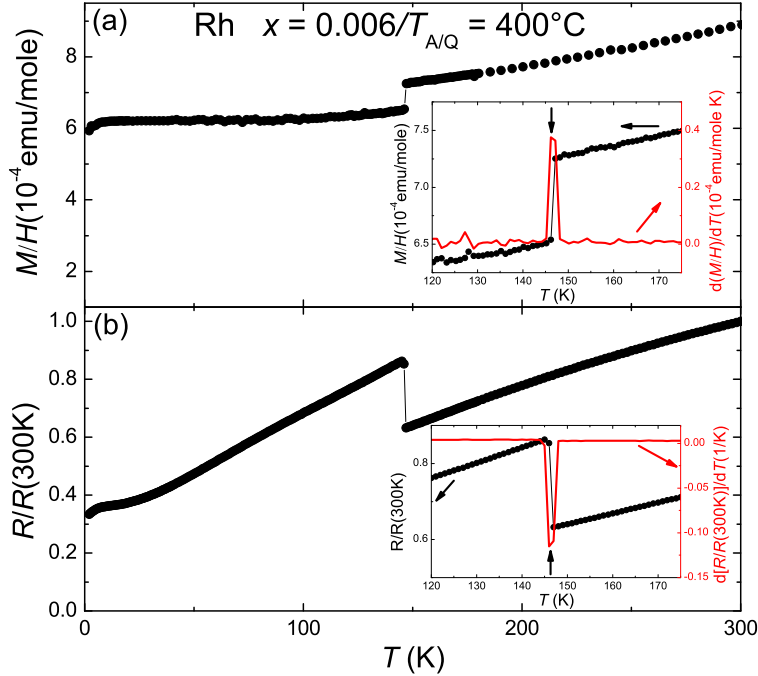}
\end{center}
\caption{(Color online) Criteria used to determine the transition temperatures of the AFM/ORTH phase transition. The data close to the transition are presented in the insets, together with the derivatives. Inferred transition temperatures are indicated by vertical arrows.}
\label{criteria1}
\end{figure}

\begin{figure}[!htbp]
\begin{center}
\includegraphics[angle=0,width=100mm]{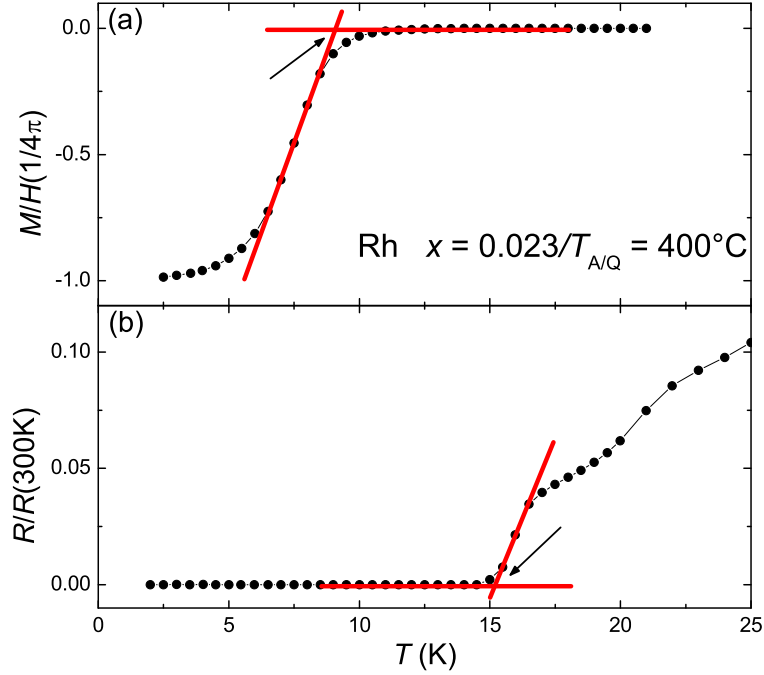}
\end{center}
\caption{(Color online) Criteria used to determine the transition temperatures of the superconducting phase transition. Inferred transition temperatures are indicated by arrows. As discussed in the text, use of resistivity data can lead to artificially high {\itshape T}$_{c}$ values due to strain-induced filamentary superconductivity.}
\label{criteria2}
\end{figure}

\begin{figure}[!htbp]
\begin{center}
\includegraphics[angle=0,width=100mm]{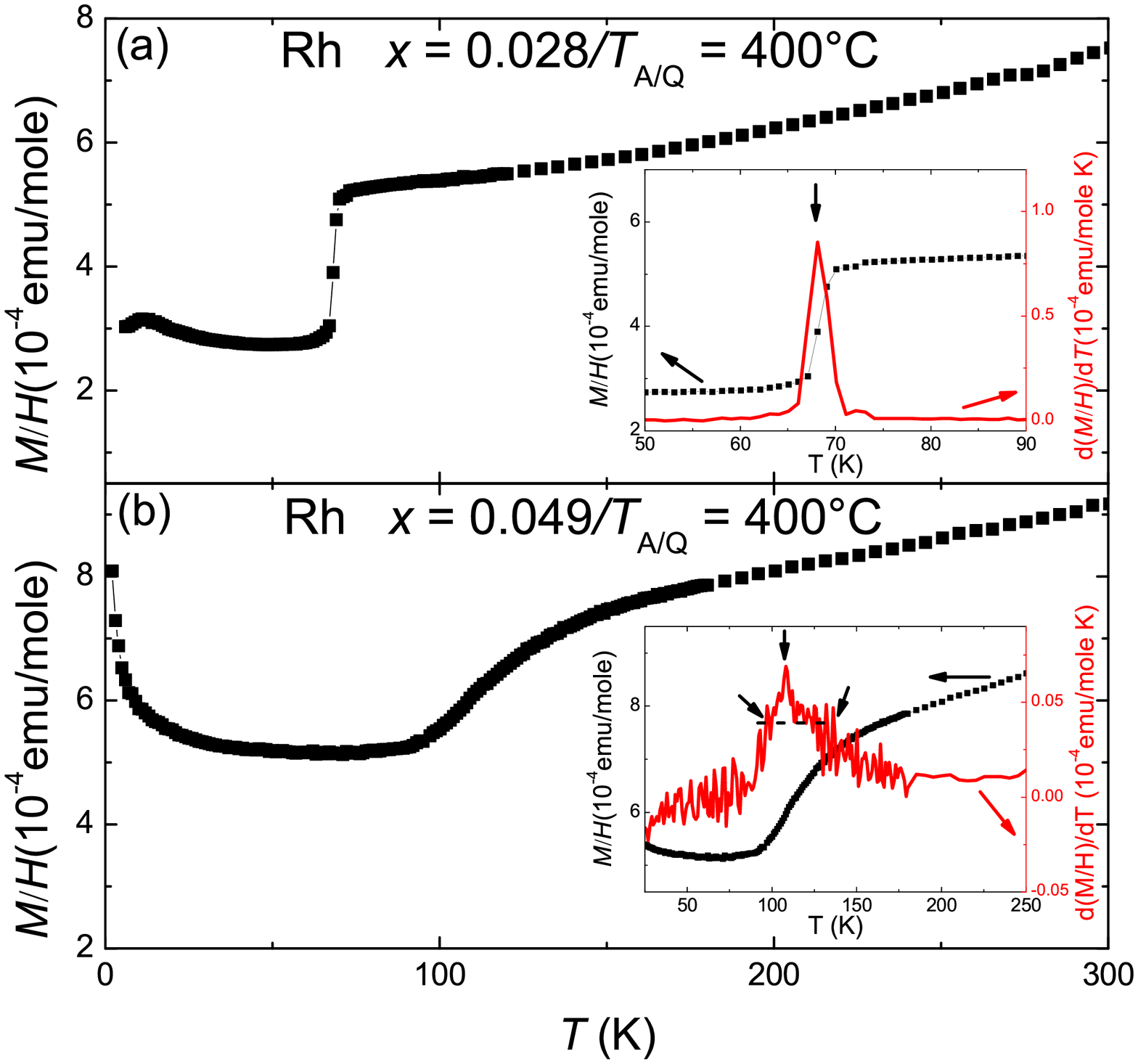}
\end{center}
\caption{(Color online) Criteria used to determine the transition temperatures of the cT phase transition shown for two different samples. The data close to the transition are presented in the insets, together with the derivatives. Inferred transition temperatures are indicated by vertical arrows.}
\label{criteria3}
\end{figure}

In order to infer phase diagrams from these thermodynamic and transport data, we need to introduce criterion for the determination of the salient transition temperatures.\cite{Ni08Ca,Goldman08,Kreyssig08,Goldman09,Canfield09Ca} The AFM/ORTH phase transition (when present) appears as a single (i.e., not split), sharp feature which is clearly identifiable in both resistance and magnetization. Figure \ref{criteria1} shows the susceptibility and resistance, as well as their temperature derivatives (insets), for a {\itshape x} = 0.006/{\itshape T}$_{A/Q}$ = 400$^\circ$C Rh-substituted sample. Clear features, including a sharp drop in susceptibility and a sharp jump in the resistance, occur upon cooling through the transition temperature. The transition temperature is even more clearly seen in the $d(M/H)/dT$ and $dR/dT$ data. For the superconducting transition, we only used an onset criterion for magnetic susceptibility (the temperature at which the maximum slope of the susceptibility extrapolates to the normal state susceptibility) to determine {\itshape T}$_{c}$. This criteria for {\itshape T}$_{c}$ is presented in Fig.\ref{criteria2}a, with an example of a {\itshape x} = 0.023/{\itshape T}$_{A/Q}$ = 400$^\circ$C Rh-substituted sample. Sometimes an offset criterion for resistance (the temperature at which the maximum slope of the resistance extrapolates to zero resistance) is also used in literature to determine {\itshape T}$_{c}$. \cite{Ni08BaCo} However, this leads to substantially higher {\itshape T}$_{c}$ than ones inferred from magnetic susceptibility in CaFe$_{2}$As$_{2}$ which, given its profound pressure and strain sensitivities, is prone to filamentary superconductivity (Fig. \ref{criteria2}b). Given that we do observe diamagnetism with zero field cooling (ZFC) susceptibility reaching 1/4$\pi$, we choose to err on the side of bulk superconductivity rather than minority phase of filamentary superconductivity.\cite{Saha12} The cT phase is induced by higher {\itshape T}$_{A/Q}$. When the cT phase transition occurs, it often leads to cracks in the resistance bar and loss of data below the transition temperature (in case the resistance bar survives upon cooling through the transition, resistance data shows downward jump and hysteresis of up to around 15~K), which is an unique fingerprint of the cT phase transition and helps us to distinguish it from AFM/ORTH phase transition. On the other hand, loss of data below the transition makes it difficult to extract an unambiguous value of the transition temperature from {\itshape R}({\itshape T}) data. Therefore only susceptibility data were used to determine the transition temperature, {\itshape T}$_{cT}$. Figure \ref{criteria3} shows the susceptibility data for two different samples. The peak in derivative of the susceptibility was employed to determine {\itshape T}$_{cT}$. Note that the peak in derivative becomes significantly broadened for high concentrations, as shown for the {\itshape x} = 0.049/{\itshape T}$_{A/Q}$ = 400$^\circ$C Rh-substituted sample. We capture the broadness of the transition by including error bars, which were defined here as the full width at half maximum of the peaks in derivatives of the susceptibility. 

\section{Results and discussion}

\subsection*{Compositional and structural determination}

\begin{figure}[!htbp]
\begin{center}
\includegraphics[angle=0,width=100mm]{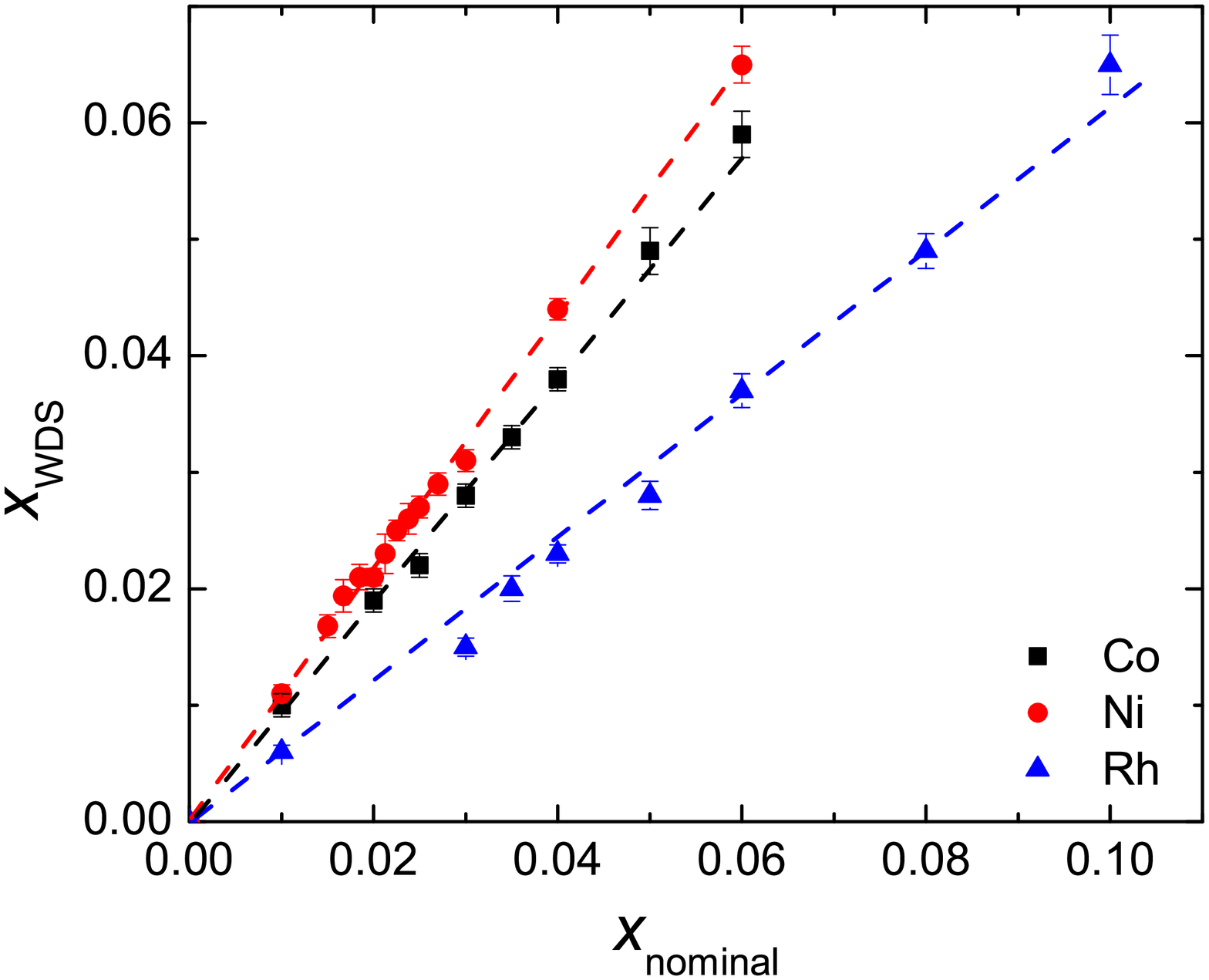}
\end{center}
\caption{(Color online) Measured Co, Ni and Rh concentration vs nominal Co, Ni and Rh concentration for the Ca(Fe$_{1-x}$T$_{x}$)$_{2}$As$_{2}$ series. Data of Co-substitution are taken from Ref. \cite{Ran12}}
\label{WDS}
\end{figure}

A summary of the WDS measurement data for both Ni- and Rh-substituted compounds is presented in Fig. \ref{WDS}. Data for the \CaCo series \cite{Ran12} are also presented for comparison. The nominal concentration versus actual concentration data for all three series can be fitted very well with straight lines, indicating a linear correlation between the measured concentration and the nominal concentration for these relatively low ({\itshape x} $<$ 0.10) substitution levels. Whereas the slope for \CaCo and \CaNi are close to 1 (0.96 $\pm$ 0.01 and 1.09 $\pm$ 0.01 respectively), it is only 0.61 $\pm$ 0.01 for Ca(Fe$_{1-x}$Rh$_{x}$)$_{2}$As$_{2}$. The error bars are taken as twice of the standard deviation determined from the 12 WDS measurements on each sample, and are no more than 0.003, demonstrating relative homogeneity of the substituted samples studied here. In the following, the average experimentally determined {\itshape x} values, {\itshape x} = {\itshape x}$_{WDS}$, will be used to identify all the compounds rather than the nominal concentration, {\itshape x}$_{nominal}$. 

\begin{figure}[!htbp]
\begin{center}
\includegraphics[angle=0,width=100mm]{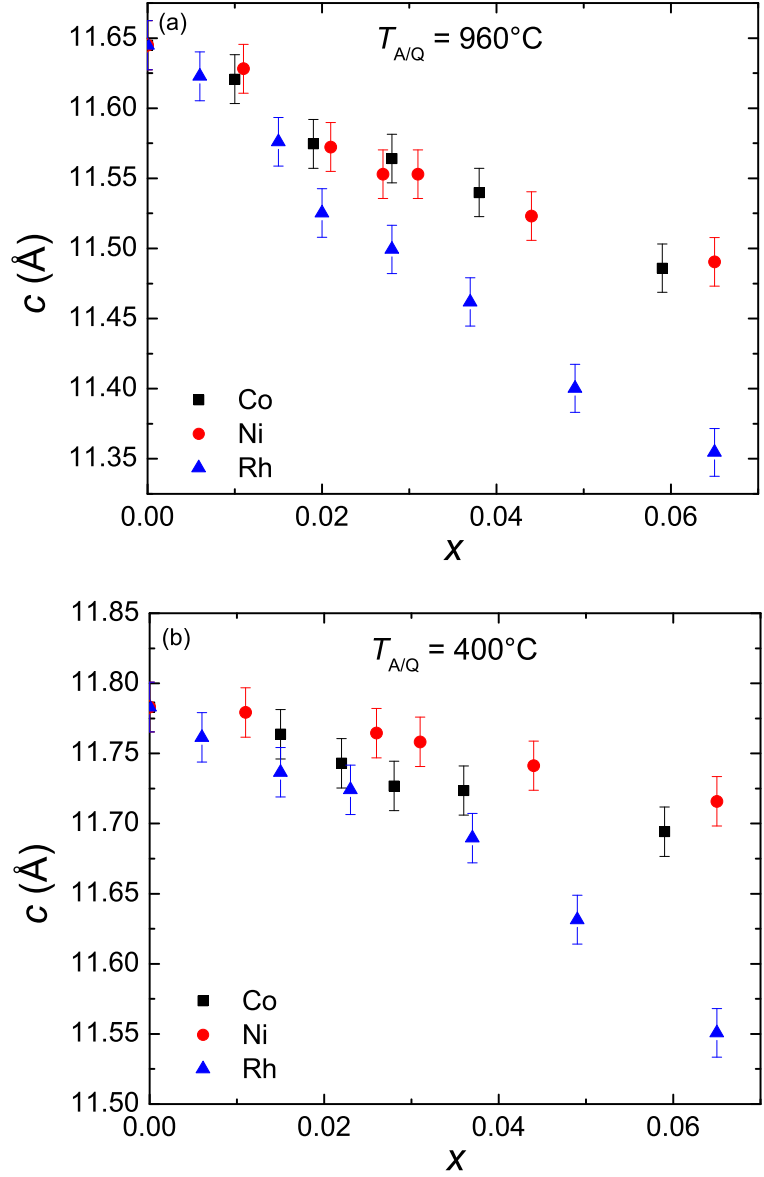}
\end{center}
\caption{(Color online) Room temperature {\itshape c}-lattice parameter for \CaNi and \CaRh series, determined via diffraction from plate like samples, as described in the Experimental Methods section, as a function of measured Ni/Rh concentration, {\itshape x} for (a) {\itshape T}$_{A/Q}$ = 960$^\circ$C samples and (b) {\itshape T}$_{A/Q}$ = 400$^\circ$C samples. For comparison, data for \CaCo samples from Ref. \onlinecite{Ran12} are also presented.}
\label{clattice}
\end{figure}

Figure~\ref{clattice} presents the {\itshape c}-lattice parameters for the {\itshape T}$_{A/Q}$ = 960$^\circ$C samples, as well as for the {\itshape T}$_{A/Q}$ = 400$^\circ$C samples, for both \CaNi and \CaRh series, determined via diffraction from the platelike samples using (002) and (008) peaks. Data for the \CaCo series \cite{Ran12} obtained in a similar way are also presented for comparison. It can be seen that in case of both the {\itshape T}$_{A/Q}$ = 960$^\circ$C samples and the {\itshape T}$_{A/Q}$ = 400$^\circ$C samples the {\itshape c}-lattice parameter is suppressed by all three transition metal substitutions. Whereas Ni-substitution suppresses the {\itshape c}-lattice parameter at roughly the same rate as Co-substitution, Rh-substitution suppresses the {\itshape c}-lattice parameter roughly twice as fast. This is similar to what has been seen for BaFe$_{2}$As$_{2}$.\cite{Ni09TM,Canfield09TM,Ni10TM,Canfield10} However, in BaFe$_{2}$As$_{2}$ this difference in the suppression of the {\itshape c}-lattice parameter does not seem to matter much in terms of its effect on the {\itshape T}-{\itshape x} phase diagrams, i.e. Co- and Rh-substitutions being virtually identical but differing from Ni- and Pd-substitutions (each with an extra conduction electron). Considering that the physical properties of \CaP are much more sensitive to the stress and strain than are those of BaFe$_{2}$As$_{2}$, the large suppression of the lattice parameter in \CaRh series may have a much more dramatic effect than in the case of Ba(Fe$_{1-x}$Rh$_{x}$)$_{2}$As$_{2}$.

\subsection*{Ca(Fe$_{1-x}$Ni$_{x}$)$_{2}$As$_{2}$}

\begin{figure}[!htbp]
\begin{center}
\includegraphics[angle=0,width=150mm]{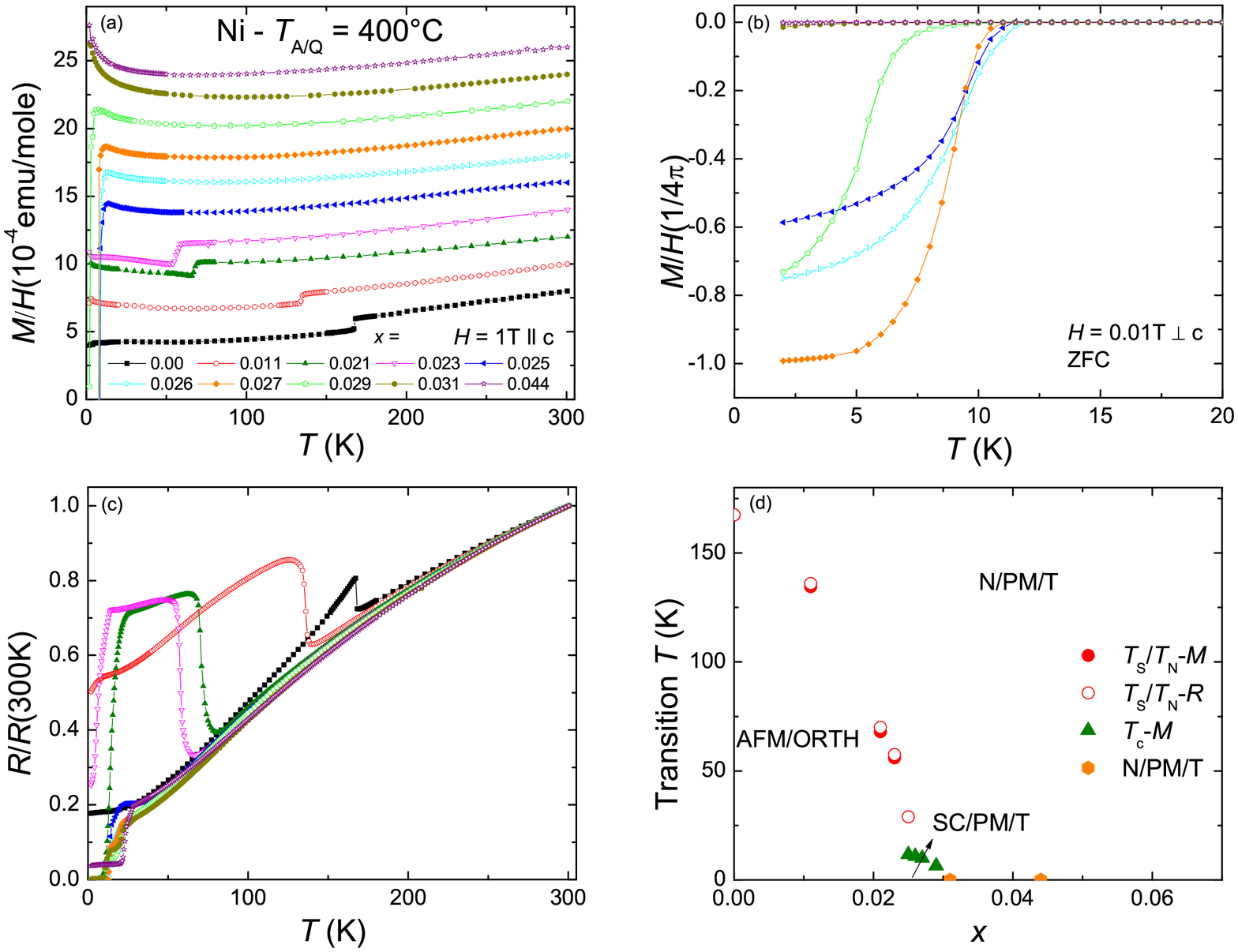}
\end{center}
\caption{(Color online) Temperature dependent (a) magnetic susceptibility with field applied parallel to the {\itshape c} axis, (b) low-field magnetic susceptibility measured upon zero field cooling (ZFC) with a field of 0.01 T applied perpendicular to the {\itshape c} axis, (c) normalized electrical resistance, and (d) phase diagram of transition temperature {\itshape T} vs Ni concentration {\itshape x} of \CaNi samples with {\itshape T}$_{A/Q}$ = 400$^\circ$C. Susceptibility data in (a) have been offset from each other by an integer multiple of 1 $\times$ 10$^{-4}$ emu/mole for clarity. Yellow symbols indicate N/PM/T state with no transition observed down to the base temperature.}
\label{Ni400C}
\end{figure}

Figure~\ref{Ni400C} presents the temperature dependent magnetic susceptibility and resistance data from \CaNi samples with {\itshape T}$_{A/Q}$ = 400$^\circ$C. The {\itshape x} = 0, parent compound, shows AFM/ORTH phase transition at around 166~K as indicated by the sharp drop in susceptibility and upward jump in resistance upon cooling. The anomalies in both susceptibility and resistance are suppressed with increasing Ni substitution level, down to 55~K, at {\itshape x} = 0.023.  For higher Ni concentrations, the AFM/ORTH phase transition is suppressed completely and the SC/PM/T phase is stabilized. At {\itshape x} = 0.025, low field susceptibility shows that the screening is around 60$\%$ of 1/4$\pi$ at 2~K as shown in Fig. ~\ref{Ni400C}b. At {\itshape x} = 0.027, the screening increases to 100$\%$ of 1/4$\pi$. Above {\itshape x} = 0.027, increasing Ni concentration suppresses {\itshape T}$_{c}$ and the screening. For {\itshape x} $\geqslant$ 0.031, the screening is suppressed to zero and the system is in a N/PM/T ground state which is neither antiferromagnetic nor superconducting.  

Based on these thermodynamic and transport measurements, we can construct a phase diagram of transition temperature versus Ni concentration for this {\itshape T}$_{A/Q}$ (Fig. \ref{Ni400C}d). The AFM/ORTH phase transition is suppressed with initial Ni-substitution and the phase line terminates at around {\itshape x} = 0.025, where the SC/PM/T phase emerges. {\itshape T}$_{c}$ is suppressed by further increasing Ni concentration. Bulk superconductivity, as indicated by low field susceptibility, is suppressed completely for {\itshape x} $\geqslant$ 0.031. No clear evidence of the coexistence of the AFM/ORTH and the SC/PM/T phases is observed and no splitting of the signature of the magnetic and structure phase transition is observed. 

In a similar manner, we constructed {\itshape T}-{\itshape x} phase diagrams for other {\itshape T}$_{A/Q}$ values as presented in Fig.~\ref{NiTxTT}a and b (see Appendix for corresponding magnetic susceptibility and resistance data). {\itshape T}$_{A/Q}$ = 500$^\circ$C leads to similar phase diagram, but, both the AFM/ORTH and the SC/PM/T phase regions are reduced. This is consistent with the fact that increasing {\itshape T}$_{A/Q}$ has the same effect as increasing pressure as shown for pure and Co-substituted compounds in our previous work. \cite{Ran11,Ran12,Gati12} For {\itshape T}$_{A/Q}$ = 960$^\circ$C, \CaP transforms into a cT state at low temperature. \cite{Ran11,Ran12} As {\itshape x} is increased in the \CaNi series the transition temperature of this cT phase is gradually suppressed. This is in contrast to the \CaCo series, where this phase transition occurs at roughly the same temperature throughout the whole substitution level in our study range.  Around {\itshape x} = 0.043, the signature of transition is broadened and, as discussed in the experimental section, the large error bar is meant to represent this. 

\begin{figure}[!htbp]
\begin{center}
\includegraphics[angle=0,width=150mm]{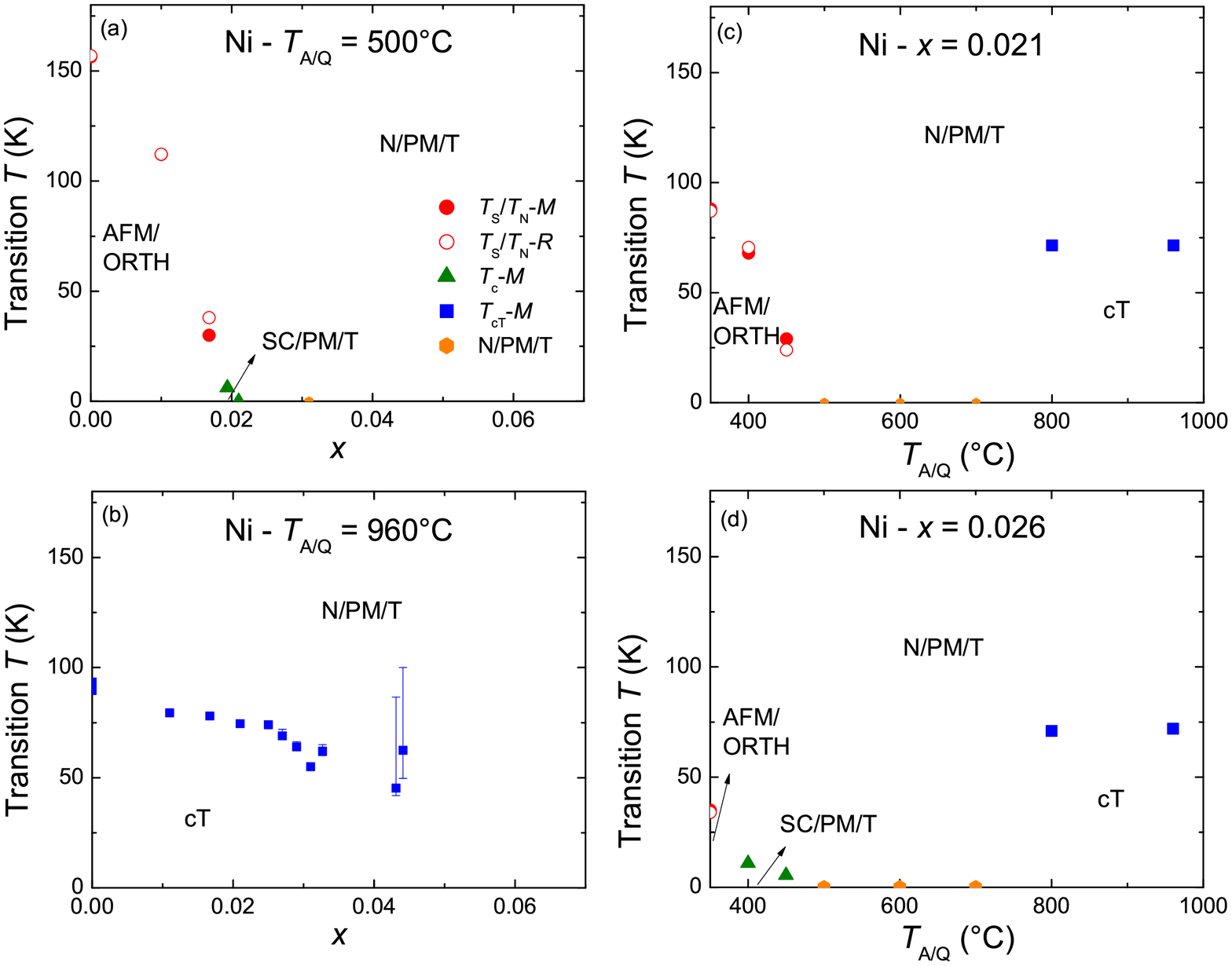}
\end{center}
\caption{(Color online) phase diagram of (a) transition temperature {\itshape T} vs Ni concentration {\itshape x} of \CaNi samples with {\itshape T}$_{A/Q}$ = 500$^\circ$C, (b) transition temperature {\itshape T} vs Ni concentration {\itshape x} of \CaNi samples with {\itshape T}$_{A/Q}$ = 960$^\circ$C, (c) transition temperature {\itshape T} vs annealing/quenching temperature {\itshape T}$_{A/Q}$ of \CaNi samples with Ni concentration {\itshape x} = 0.021, and (d) transition temperature {\itshape T} vs annealing/quenching temperature {\itshape T}$_{A/Q}$ of \CaNi samples with Ni concentration {\itshape x} = 0.026. Yellow symbols indicate N/PM/T state with no transition observed down to the base temperature.}
\label{NiTxTT}
\end{figure}
 
In order to systematically study the effect of the varying {\itshape T}$_{A/Q}$ for a given Ni substitution level, we studied {\itshape x} = 0.021 and {\itshape x} = 0.026 samples for 350$^\circ$C $\leqslant$ {\itshape T}$_{A/Q}$ $\leqslant$ 960$^\circ$C. The corresponding phase diagrams are presented in Fig.~\ref{NiTxTT}c and d. For both {\itshape x} = 0.021 and {\itshape x} = 0.026, the ground state of the \CaNi series is AFM/ORTH phase for low {\itshape T}$_{A/Q}$ ($\leqslant$ 450$^\circ$C for {\itshape x} = 0.021 and $\leqslant$ 350$^\circ$C for {\itshape x} = 0.026) and cT phase for high {\itshape T}$_{A/Q}$ ($\geqslant$ 800$^\circ$C). For intermediate values of {\itshape T}$_{A/Q}$, no bulk superconductivity (i.e., with significant screening) is observed for {\itshape x} = 0.021, whereas for {\itshape x} = 0.026, bulk superconductivity with screening of more than 70$\%$ of 1/4$\pi$ at 2~K is observed for {\itshape T}$_{A/Q}$ = 400$^\circ$C.

\begin{figure}[!htbp]
\begin{center}
\includegraphics[angle=0,width=150mm]{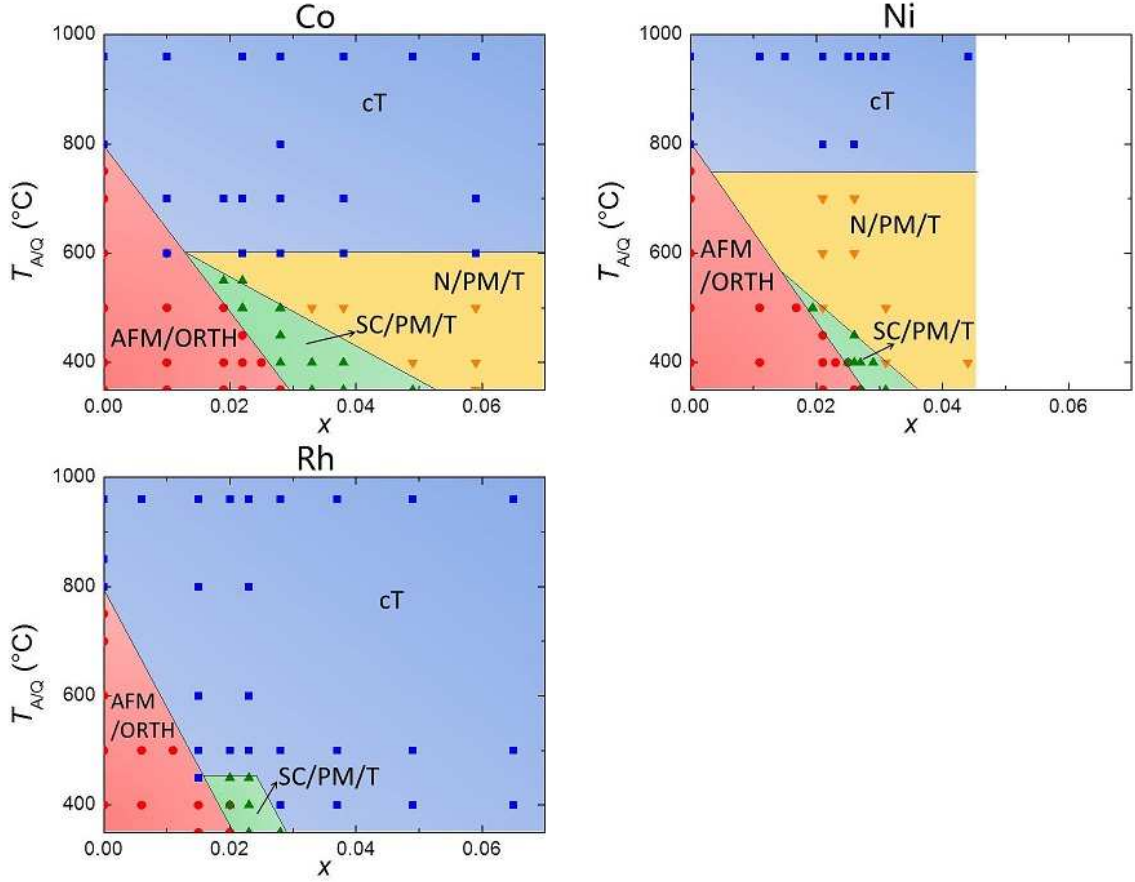}
\end{center}
\caption{(Color online) 2D phase diagrams, with transition metal concentration {\itshape x} and annealing/quenching temperature {\itshape T}$_{A/Q}$ as two independent variables, for (a) Ca(Fe$_{1-x}$Co$_{x}$)$_{2}$As$_{2}$,\cite{Ran12} (b) Ca(Fe$_{1-x}$Ni$_{x}$)$_{2}$As$_{2}$, and (c) Ca(Fe$_{1-x}$Rh$_{x}$)$_{2}$As$_{2}$. The red area delineates the conditions that lead to AFM/ORTH phase as ground state. The green area delineates the conditions that lead to SC/PM/T phase as ground state. The yellow area delineates the conditions that lead to N/PM/T phase as ground state. The blue area delineates the conditions that lead to cT phase as ground state.}
\label{2D}
\end{figure}

Since in this work we mainly focus on mapping out the relationship between possible low temperature states for various combinations of substitution level and annealing/quenching temperature, we can construct a 2D phase diagram, with transition metal concentration {\itshape x} and annealing/quenching temperature {\itshape T}$_{A/Q}$ as two independent variables, and mark the ground state with different symbols (and colors online). This phase diagram is essentially a projection of the 3D phase diagram (as in Ref. \cite{Ran12}) onto the plane of base temperature. Based on the magnetic susceptibility and resistance data, we assembled a 2D phase diagram for Ni-substitution and compare it with that of Co-substitution, as shown in Fig.~\ref{2D}. 

As seen for the \CaCo system, the \CaNi system also possesses the same salient low temperature states associated with Fe-based superconductors: AFM/ORTH, SC/PM/T, N/PM/T and cT. The AFM/ORTH region found for both Ni- and Co-substitution span essentially the same parameter space, whereas the SC/PM/T region for Ni-substitution is significantly reduced compared with Co-substitution, with maximum {\itshape x} value that supports SC/PM/T being much smaller than that for Co-substitution (for {\itshape T}$_{A/Q}$ = 350$^\circ$, the critical substitution level is roughly 3.5$\%$ for Ni and 5.5$\%$ for Co). This is consistent with what was found for Co- and Ni-substituted BaFe$_{2}$As$_{2}$,\cite{Ni09TM,Canfield09TM,Ni10TM,Canfield10} where antiferromagnetism seems to be primarily controlled by the impurity concentration {\itshape x} whereas superconductivity was more strongly influenced by extra electron count, {\itshape e}. In addition, the cT phase region for Ni-substituted \CaP is also reduced. The cT phase is only stabilized for {\itshape T}$_{A/Q}$ $\geqslant$ 800$^\circ$C as opposed to {\itshape T}$_{A/Q}$ $\geqslant$ 500$^\circ$C for Co-substitution. 

\subsection*{Ca(Fe$_{1-x}$Rh$_{x}$)$_{2}$As$_{2}$}

Rh-substitution brings the same nominal amount of extra electrons as Co-substitution does and, despite the generic difference between 3d-shell and 4d-shell electrons as well as a much more rapid change in the {\itshape c} lattice parameter. In the case of Ba(Fe$_{1-x}$Rh$_{x}$)$_{2}$As$_{2}$, Rh-substitution leads to a virtually identical {\itshape T}-{\itshape x} phase diagram as found for Ba(Fe$_{1-x}$Co$_{x}$)$_{2}$As$_{2}$. Given that {\CaP} is much more sensitive to the pressure and strain than BaFe$_{2}$As$_{2}$, the different steric effects may well lead to differences in the {\itshape T}$_{A/Q}$-{\itshape x} phase diagrams in the case Ca(Fe$_{1-x}$Co$_{x}$)$_{2}$As$_{2}$ of Ca(Fe$_{1-x}$Rh$_{x}$)$_{2}$As$_{2}$.

\begin{figure}[!htbp]
\begin{center}
\includegraphics[angle=0,width=150mm]{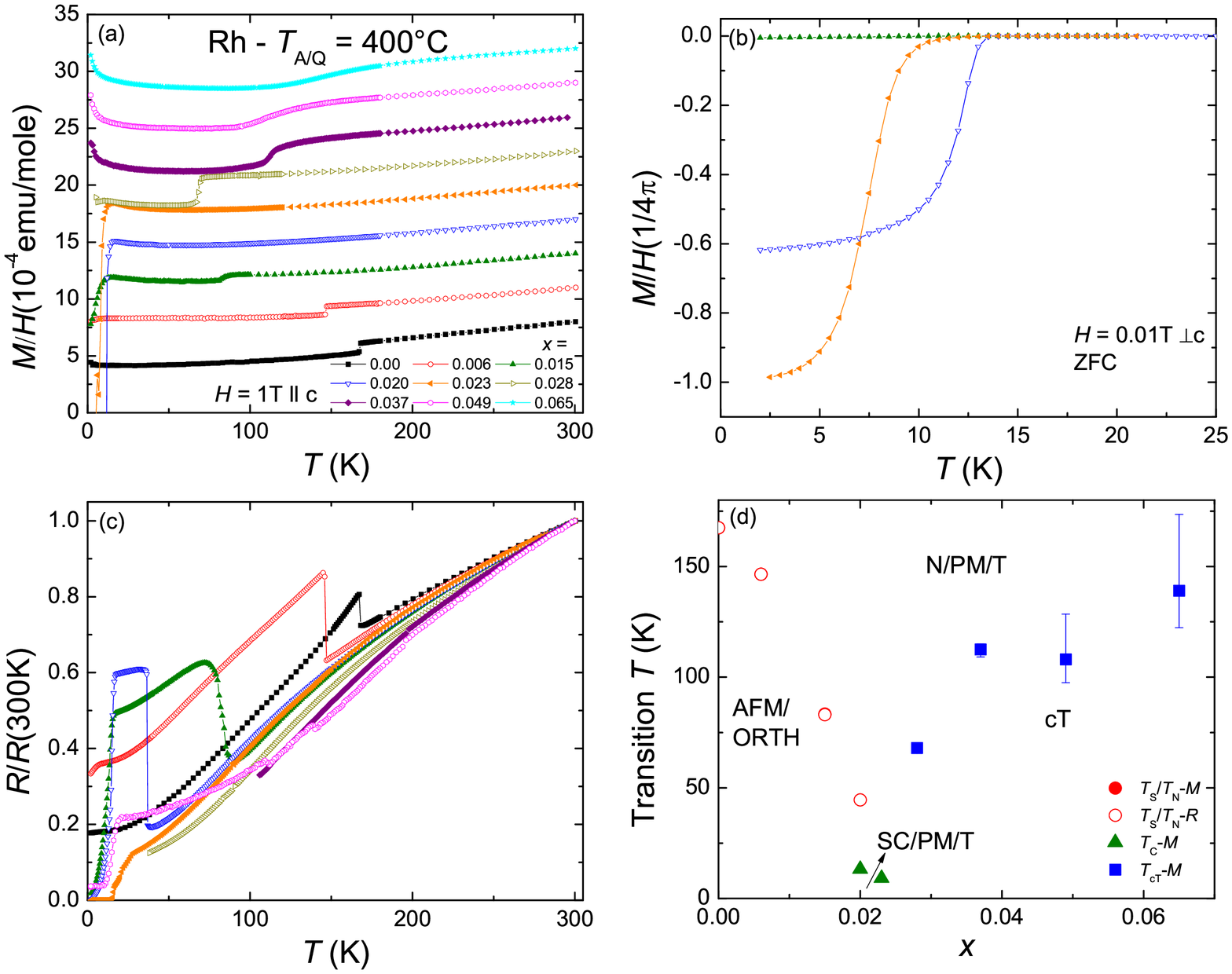}
\end{center}
\caption{(Color online) Temperature dependent (a) magnetic susceptibility with field applied parallel to the {\itshape c} axis, (b) low-field magnetic susceptibility measured upon ZFC with a field of 0.01 T applied perpendicular to the {\itshape c} axis, (c) normalized electrical resistance, and (d) phase diagram of transition temperature {\itshape T} vs Rh concentration {\itshape x} of \CaRh samples with {\itshape T}$_{A/Q}$ = 400$^\circ$C. Susceptibility data in (a) have been offset from each other by an integer multiple of 3 $\times$ 10$^{-4}$ emu/mole for clarity.}
\label{Rh400C}
\end{figure}

Figures~\ref{Rh400C}a to c present the magnetic susceptibility and resistance data for \CaRh compounds with {\itshape T}$_{A/Q}$ = 400$^\circ$. Rh-substitution initially suppresses the AFM/ORTH transition to below 50~K by {\itshape x} = 0.02. Bulk superconductivity is observed in a small region of {\itshape x} value, as shown by screening in low field susceptibility (\ref{Rh400C}b). Unlike the cases of Co- or Ni-substitution, both of which which have a region of {\itshape T}$_{A/Q}$-{\itshape x} values that lead to a N/PM/T ground state without bulk superconductivity (\ref{2D}a and b), Rh-substitution stabilizes the cT state much more rapidly, precluding any N/PM/T phase and abruptly terminating its SC/PM/T region in a manner similar to what is seen for application of hydrostatic pressure to superconducting samples of Ca(Fe$_{1-x}$Co$_{x}$)$_{2}$As$_{2}$.\cite{Gati12} Given that previous work showed that both cT and AFM/ORTH phases are much more sensitive to changes in the {\itshape c}-axis than to changes in the {\itshape ab}-axis,\cite{Sergey13} this can be understood based on the fact that Rh-substitution suppresses {\itshape c}-lattice parameter more rapidly than either Co- or Ni-substitution. The cT phase line starts near 70~K at {\itshape x} = 0.028 and reaches 140~K at {\itshape x} = 0.065, where the transition becomes broadened as also seen for the {\itshape T}$_{A/Q}$ = 960$^\circ$C, high substitution levels. The three low temperature states can be seen in the phase diagram presented in Fig.~\ref{Rh400C}d. Note that at {\itshape x} = 0.02, low field magnetic susceptibility shows superconducting signal with screening of more than 60$\%$ of 1/4$\pi$, whereas resistance data (which was taken on the same piece of sample) shows upward turning upon cooling indicating AFM/ORTH transition. Given that {\itshape x} = 0.02 is at the phase boundary, it is very likely that part of the sample transforms into SC/PM/T phase and the other part of the sample transforms into AFM/ORTH phase. The other possibility is the coexistence of superconductivity and antiferromagnetism. This scenario is unlikely because the AFM/ORTH phase transition in this compound remains quite first order even though it is suppressed to around 50~K, as indicated by the sharpness of the resistive signature of the transition. Therefore, there would not be enough magnetic fluctuations, which are vital for the emergence of the unconventional superconductivity in the iron pnictides according to the current theories,\cite{Mazin08,Fernandes10} as well as our earlier experiments on Ca(Fe$_{1-x}$Co$_{x}$)$_{2}$As$_{2}$,\cite{Ran12,Gati12} to support superconductivity in the AFM/ORTH state.

\begin{figure}[!htbp]
\begin{center}
\includegraphics[angle=0,width=150mm]{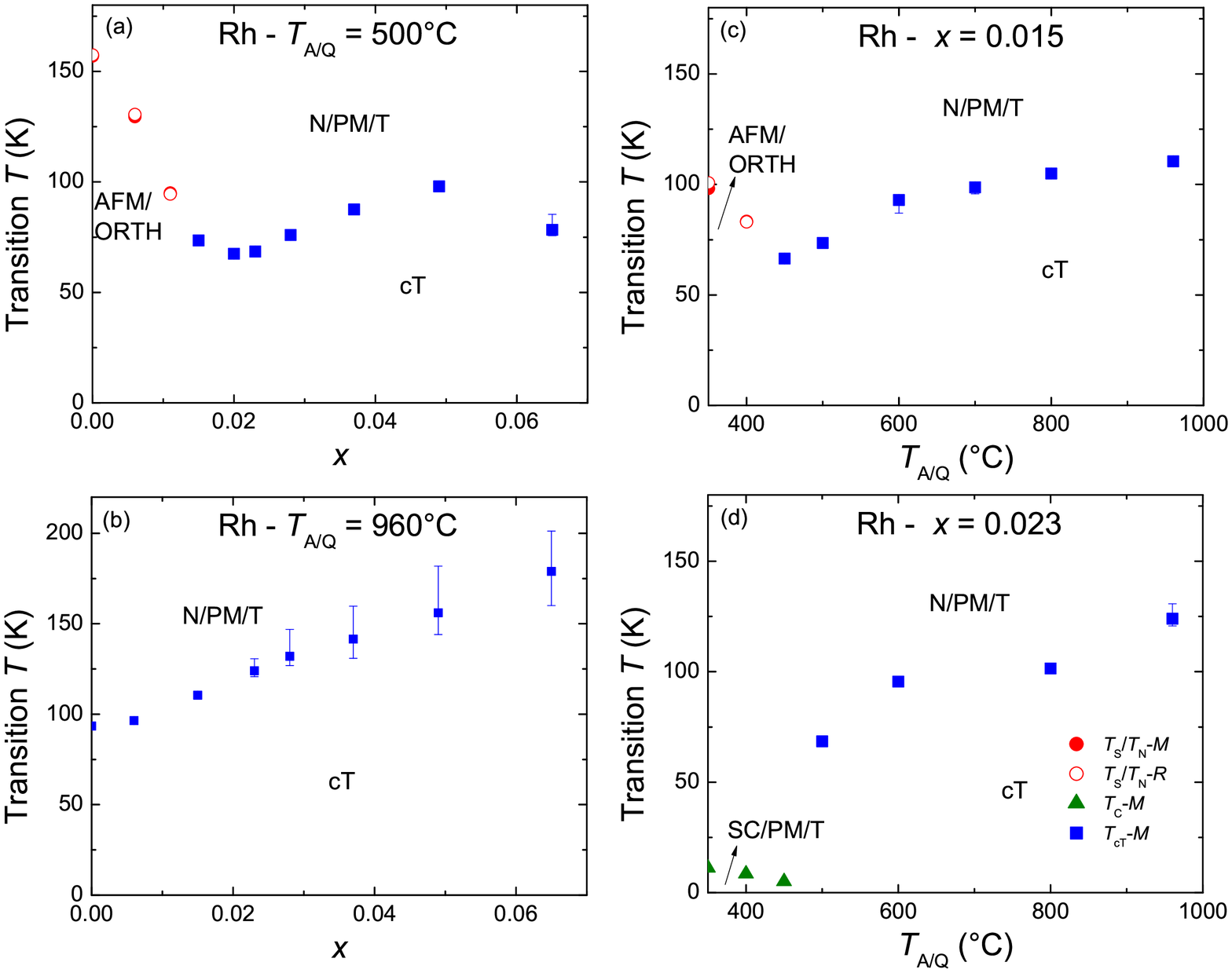}
\end{center}
\caption{(Color online)  phase diagram of (a) transition temperature {\itshape T} vs Rh concentration {\itshape x} of \CaRh samples with {\itshape T}$_{A/Q}$ = 500$^\circ$C, (b) transition temperature {\itshape T} vs Rh concentration {\itshape x} of \CaRh samples with {\itshape T}$_{A/Q}$ = 960$^\circ$C, (c) transition temperature {\itshape T} vs annealing/quenching temperature {\itshape T}$_{A/Q}$ of \CaRh samples with Rh concentration {\itshape x} = 0.015, and (d) transition temperature {\itshape T} vs annealing/quenching temperature {\itshape T}$_{A/Q}$ of \CaRh samples with Rh concentration {\itshape x} = 0.023.}
\label{RhTxTT}
\end{figure}

Figure \ref{RhTxTT} presents {\itshape T}-{\itshape x} phase diagrams for different annealing/quenching temperatures and {\itshape T}-{\itshape T}$_{A/Q}$ phase diagrams for different Rh concentrations. Similar to what we did for Ni-substitution, we assembled these data and constructed a 2D phase diagram for the base-temperature states of the \CaRh system, as presented in Fig. \ref{2D}. The {\itshape T}$_{A/Q}$-{\itshape x}, 2D phase diagram of Rh-substitution is significantly different from that of Co-substitution. This is in contrast to the case of Ba(Fe$_{1-x}$T$_{x}$)$_{2}$As$_{2}$, where the phase diagrams for Co- and Rh-substitutions are almost identical.\cite{Ni09TM,Canfield09TM,Ni10TM,Canfield10} In the case of \CaRh the AFM/ORTH phase is suppressed faster than it is for Co-substitution and the cT phase is much more pervasive in case of Rh-substitution, appearing for all annealing/quenching temperatures for substitutions level above 3$\%$. Both of these changes can be understood based on the fact that Rh-substitution suppresses the {\itshape c}-lattice parameter more rapidly than Co (or Ni) substitution. A consequence of the enhanced stabilization of the cT phase for low {\itshape T}$_{A/Q}$ values is (i) the complete absence on the N/PM/T phase and (ii) the SC/PM/T region for Rh-substitution is substantially shrunk, or truncated, compared with that for Co-substitution. Given that (i) current theories and experiments indicate that the spin fluctuations play an important role for the appearance of unconventional superconductivity in the iron pnictides; and (ii) spin fluctuations are completely suppressed in the cT phase in CaFe$_{2}$As$_{2}$, \cite{Yildirim09,Ran11,Pratt09,Soh13,Dhaka14,Furukawa14} it is likely that the superconductivity is limited by the pervasive cT phase in the \CaRh system.

\subsection*{Critical {\itshape c}-lattice parameter}

\begin{figure}[!htbp]
\begin{center}
\includegraphics[angle=0,width=100mm]{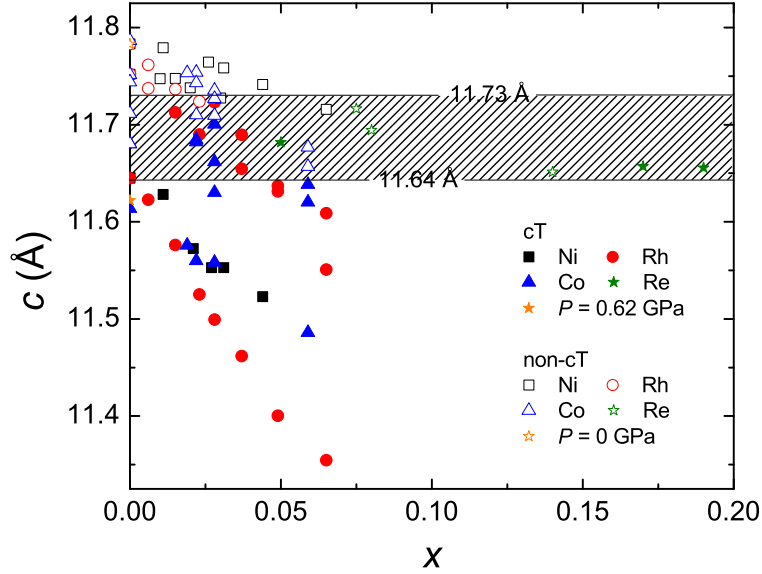}
\end{center}
\caption{(Color online) {\itshape c}-lattice parameter versus substitution level of all three substitutions. Data of rare earth substitution from Ref.\cite{Saha12} and data for Sn-grown pure CaFe$_{2}$As$_{2}$ under pressure from Ref.\cite{Kreyssig08} are also included for comparison.}
\label{criticalc}
\end{figure}

The cT phase transition is driven by an increasing overlap of interlayer As orbitals.\cite{Yildirim09} While it was suggested that As-As interlayer separation appears to be the key parameter controlling the volume collapse when comparing members of the ThCr$_{2}$Si$_{2}$ structure,\cite{Saha12,Hoffmann85} it is conceivable that, for substitutions to CaFe$_{2}$As$_{2}$, there might also be a critical room temperature {\itshape c}-lattice parameter value. Given that As-As interlayer separation is hard to measure, a critical room temperature {\itshape c}-lattice parameter value can give an easy evaluation of whether the system will transform into the cT phase or not. In order to assess the extent to which such a critical value can be inferred, we plotted the {\itshape c}-lattice parameter versus substitution level, {\itshape x}, for all three substitutions with various {\itshape T}$_{A/Q}$, as shown in Fig. \ref{criticalc}. Rare earth substitution data from literature,\cite{Saha12} as well as data for Sn-grown, {\itshape x} = 0 CaFe$_{2}$As$_{2}$ under pressure,\cite{Kreyssig08} are also presented for comparison. It can be seen that the room temperature {\itshape c}-lattice parameter can be divided into three regions: (i) below 11.64 $\AA$, where all the samples transform into cT phase at low temperature; (ii) above 11.73 $\AA$, where all the samples have non-cT phase as low temperature ground state; (iii) between 11.64 $\AA$ and 11.73 $\AA$, where details, such as temperature dependence of thermal contraction, amount of internal strain, specific type of substitution, etc., become important for determining the low temperature structural state. Note that all the rare earth substituted samples fall into the last category which is consistent with the fact that detailed As-As interlayer separation determines the ground state. The As-As interlayer separation of the Ce-substituted samples with {\itshape x} = 0.16, when extrapolated to base temperature assuming a constant temperature dependence, is just above the claimed critical value.\cite{Saha12} On the other hand, the room temperature {\itshape c}-lattice parameter, 11.65 $\AA$, is also on the edge of the last region, showing good agreement with the criteria of As-As interlayer separation. The data for Sn-grown, {\itshape x} = 0 sample also fit our criteria very well. Under the ambient pressure, the room temperature {\itshape c}-lattice parameter falls into the second category with the low temperature state being a AFM/ORTH phase, whereas under the pressure of 0.62 GPa, the {\itshape c}-lattice parameter, when extrapolated to room temperature, falls into the first category with the low temperature state being a cT phase.   

\subsection*{Annealing time dependence}

\begin{figure}[!htbp]
\begin{center}
\includegraphics[angle=0,width=100mm]{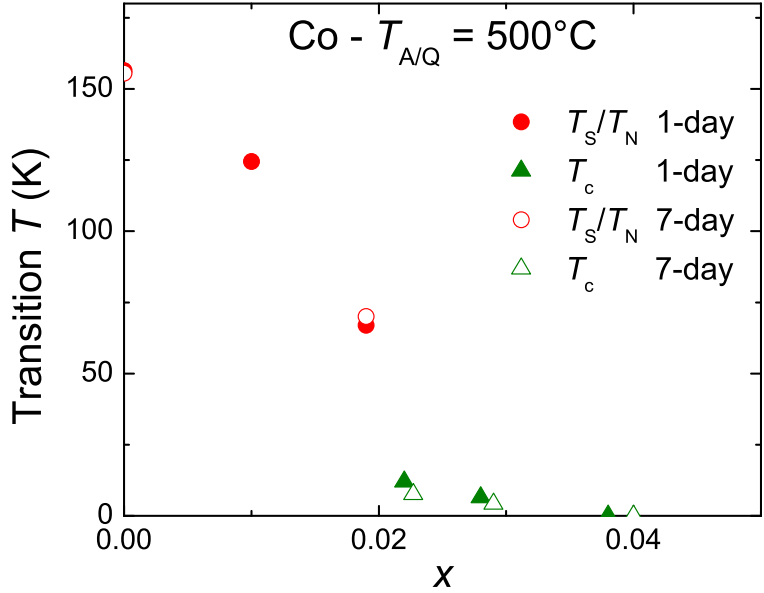}
\end{center}
\caption{(Color online) Phase diagrams of transition temperature {\itshape T} vs Co concentration {\itshape x} assembled from magnetic susceptibility data,  for \CaCo samples with {\itshape T}$_{A/Q}$ = 500$^\circ$C.\cite{Ran12} Filled symbols are inferred from data from samples with 1-day annealing and open symbols are inferred from data from samples with 7-day annealing.}
\label{Co500C}
\end{figure}

For earlier work on both pure \CaP \cite{Ran11} and Ca(Fe$_{1-x}$Co$_{x}$)$_{2}$As$_{2}$,\cite{Ran12} we performed systematic studies of effects of annealing time for various {\itshape T}$_{A/Q}$ and showed that the effects of annealing were established rather quickly (t $<$ 24 h) for {\itshape T}$_{A/Q}$ of interest. In addition, for both pure \CaP and \CaCo we found that longer annealing time did not significantly change the {\itshape T}-{\itshape T}$_{A/Q}$ phase diagrams indicating that there was only one salient annealing process with a single characteristic time. As an example, virtually identical phase diagrams of \CaCo for {\itshape T}$_{A/Q}$ = 500$^\circ$, assembled from two different sets of data, 1-day anneal and 7-day anneal are presented in Fig.~\ref{Co500C}.

\begin{figure}[!htbp]
\begin{center}
\includegraphics[angle=0,width=100mm]{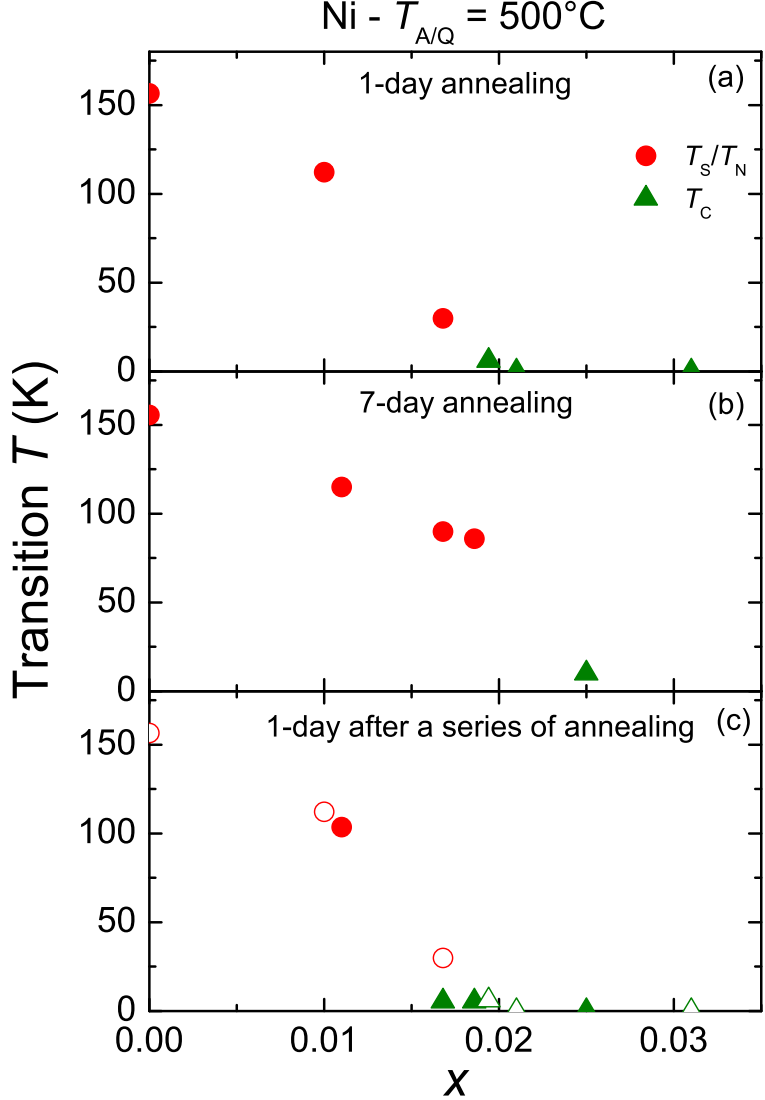}
\end{center}
\caption{(Color online) Phase diagrams of transition temperature {\itshape T} vs Ni concentration {\itshape x} assembled from magnetic susceptibility data, for \CaNi samples with {\itshape T}$_{A/Q}$ = 500$^\circ$C. (a) 1-day annealing, (b) 7-day annealing, (c) 1-day annealing after a series of annealing described in the text. For comparison, data in (a) are repeated in (c) with open symbols.}
\label{Ni500C7dPD}
\end{figure}

Ni- and Rh-substitutions appear to be different. Although 1-day annealing gives familiar phase diagrams, they change with longer annealing times. Figure \ref{Ni500C7dPD} presents the phase diagrams for Ni-substitution for {\itshape T}$_{A/Q}$ = 500$^\circ$C with different annealing time sequences. As can be seen, for samples annealed for seven days, the AFM/ORTH phase transition is suppressed more slowly and the SC/PM/T phase is only stabilized for a slightly higher Ni concentration level. However, the reproducibility with respect to annealing/quenching history seems to be preserved. We took these 7-day, 500$^\circ$C annealed samples, resealed them, annealed/quenched at 800 $^\circ$C trying to bring the samples back to a state that is close to {\itshape T}$_{A/Q}$ = 960$^\circ$C samples, and then annealed again at 500$^\circ$C for one day and quenched. After this series of annealing, the {\itshape T}-{\itshape x} phase diagram is similar to that is seen for the initial 1-day annealing, indicating that whatever process is taking place over this longer time scale, it is reversible. These data imply that (i) there is more than one salient annealing time, but that (ii) there is clear reversibility and reproducibility. 

\begin{figure}[!htbp]
\begin{center}
\includegraphics[angle=0,width=100mm]{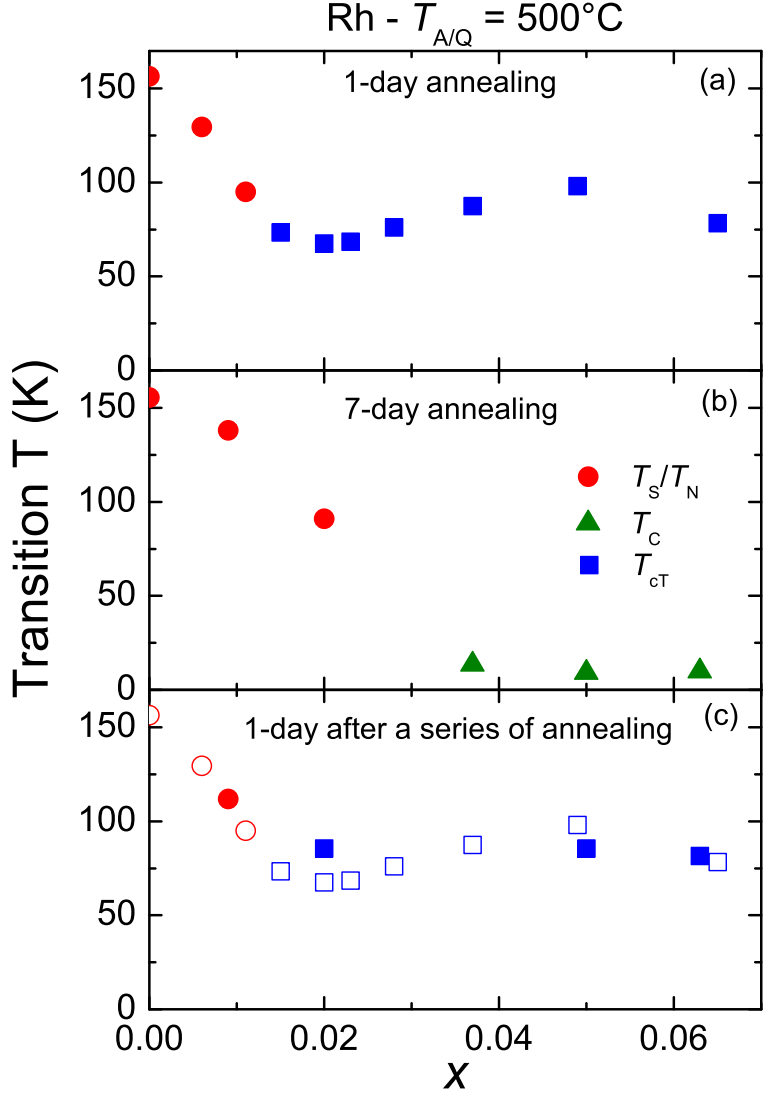}
\end{center}
\caption{(Color online) Phase diagrams of transition temperature {\itshape T} vs Rh concentration {\itshape x} assembled from magnetic susceptibility data, for \CaRh samples with {\itshape T}$_{A/Q}$ = 500$^\circ$C. (a) 1-day annealing, (b) 7-day annealing, (c) 1-day annealing after a series of annealing described in the text. For comparison, data in (a) are repeated in (c) with open symbols.}
\label{Rh500C7dPD}
\end{figure}

Even larger effects of a longer annealing time are observed for Rh-substitution as shown in Fig.~\ref{Rh500C7dPD}. It can be seen that the AFM/ORTH phase transition is initially suppressed more slowly for the 7-day annealed/quenched samples than for the 1-day annealed/quenched samples. In addition, the SC/PM/T ground state is stabilized at low temperature for the 7-day annealed/quenched samples with substitution level of 3.7$\%$ and higher. This is in stark contrast to what has been seen for 1-day annealed/quenched \CaRh compounds, where the cT phase is found for high substitution level and no superconductivity is revealed. Again we resealed these 7-day annealed samples, annealed/quenched at 800 $^\circ$C, and then annealed at 500$^\circ$C for one day and quenched. As seen for Ni-substituted samples, after this series of annealing, the initial \textquotedblleft 1-day anneal" phase diagram is recovered, illustrating clear reversibility and reproducibility.   

\begin{figure}[!htbp]
\begin{center}
\includegraphics[angle=0,width=100mm]{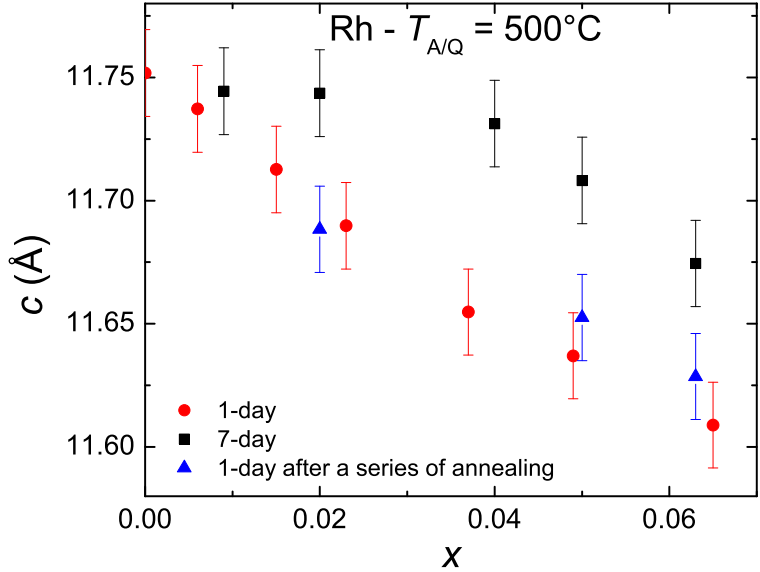}
\end{center}
\caption{(Color online) Room temperature {\itshape c}-lattice parameter of \CaRh samples with {\itshape T}$_{A/Q}$ = 500$^\circ$C as a function of measured Rh concentration, {\itshape x}, for 1-day annealing (red), 7-day annealing (black) and 1-day annealing after a series of annealing described in the text (blue). }
\label{Rh500Clattice}
\end{figure}

The clear difference between effects of 1-day and 7-day annealing, as well as the clear reversibility and reproducibility, can also be seen in the {\itshape c}-lattice parameter data from the \CaRh system as presented in Fig.~\ref{Rh500Clattice}. The {\itshape c}-lattice parameter is suppressed by Rh-substitution much less rapidly for 7-day, 500$^\circ$C annealed/quenched samples, than for 1-day, 500$^\circ$C annealed/quenched samples. After a series of further thermal treatment, we could bring it back to the behavior similar to what is seen for 1-day, 500$^\circ$C annealed/quenched samples. 

The origin of this annealing time dependence of the physical properties is still unknown. One possibility is that there are two salient time scales. One timescale for the small, excess of FeAs associated with the CaFe$_{2}$As$_{2}$ width of formation to go in and out of the CaFe$_{2}$As$_{2}$ matrix, as we proposed based on our {\itshape T}-{\itshape T}$_{A/Q}$ phase diagram and TEM results.\cite{Ran11,Ran12} Another timescale for some Fe/Ni (and Fe/Rh) segregation. Note that this is only speculation but would fit the data. As we change annealing times around the second time scale, we would change the Rh/Ni (or RhAs/NiAs) content and therefore change the phase diagram as well as the {\itshape c}-lattice parameter in a reversible manner. The fact that Co-substitution does not show the same annealing time dependence raises the question of what the differences between solubility of Co and Rh/Ni or CoAs and RhAs/NiAs in the \CaP matrix are. More detailed microscopic study, such as high resolution TEM, will be needed to provide further insight into this issue.

\section{Summary}

We report systematic studies of the combined effects of annealing/quenching temperature and Ni/Rh-substitution on the physical properties of CaFe$_{2}$As$_{2}$. We constructed two-dimensional phase diagrams for the low-temperature states for both systems to map out the relations between possible ground states and then compared with that of Co-substitution. Ni-substitution, which brings one more extra electron per substituted atom and suppresses the {\itshape c}-lattice parameter at roughly the same rate as Co-substitution, leads to similar changes in the Ca(Fe$_{1-x}$Ni$_{x}$)$_{2}$As$_{2}$ phase diagram as were seen when comparing the Ba(Fe$_{1-x}$Co$_{x}$)$_{2}$As$_{2}$ and Ba(Fe$_{1-x}$Ni$_{x}$)$_{2}$As$_{2}$ phase diagrams: similar suppression of the AFM/ORTH phase but a more rapid suppression of the SC/PM/T phase for Ni-substitution. On the other hand, Rh-substitution, which brings the same amount of extra electrons but suppresses the {\itshape c}-lattice parameter more rapidly that Co-substitution, has a very different phase diagram from that of Ca(Fe$_{1-x}$Co$_{x}$)$_{2}$As$_{2}$: Rh-substitution suppresses the AFM/ORTH phase more rapidly than Co-substitution, but more dramatically, the cT phase is stabilized over a much greater region of the {\itshape x}-{\itshape T}$_{A/Q}$ phase space, truncating the SC/PM/T region. In addition to the differences in phase diagrams, we also found different behavior in both systems related to annealing time compared to Co-substitution. We propose that for Ni- and Rh-substitution, there is a second, reversible process taking place on a longer time scale, but at the current time we do not know its microscopic origin.

\begin{acknowledgements}
S.Ran acknowledges Anton Jesche for help on the x-ray measurement. Authors acknowledge Matthew Kramer and Lin Zhou for usefully discussions. Work at the Ames Laboratory was supported by the Department of Energy, Basic Energy Sciences, Division of Materials Sciences and Engineering under Contract No. DE-AC02-07CH11358. S.L.B. acknowledges partial support from the State of Iowa through Iowa State University.
\end{acknowledgements}

\clearpage

\appendix*
\section*{Appendix}

This appendix presents the magnetic susceptibility and resistance data for both Ni- and Rh-substitutions that were used to construct phase diagrams presented and discussed in the main text. For as grown samples (which were quenched from 960$^\circ$C), due to the violent structure phase transition associated with the cT phase transition, the resistance measurements suffer from cracking and contact problems. Therefore only magnetic susceptibility data are presented.

\subsection*{Ni-substitution {\itshape T}$_{A/Q}$ = 500$^\circ$C}

\begin{figure}[!htbp]
\begin{center}
\includegraphics[angle=0,width=150mm]{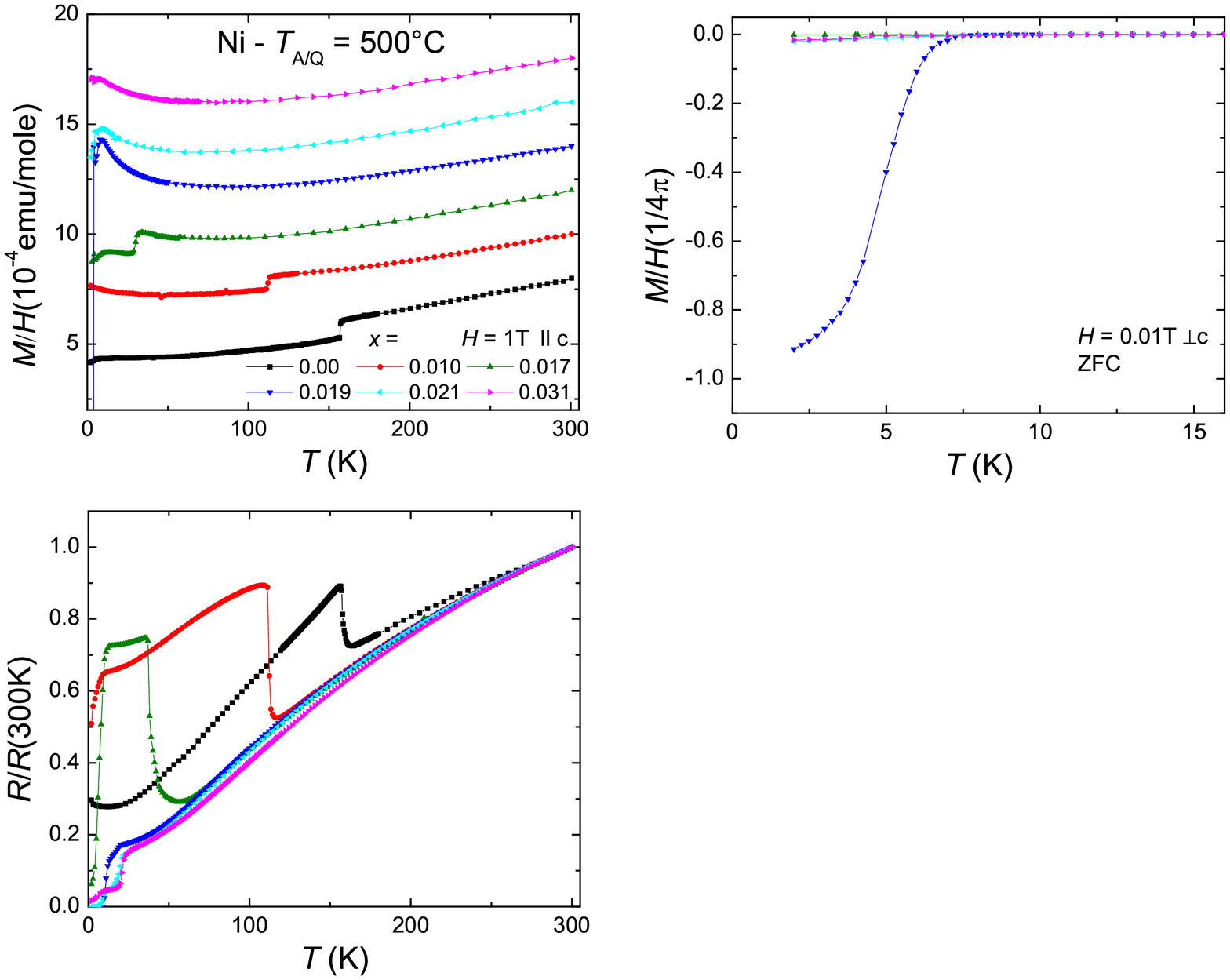}
\end{center}
\caption{(Color online) Temperature dependent (a) magnetic susceptibility with field applied parallel to the {\itshape c} axis, (b) low-field magnetic susceptibility measured upon ZFC with a field of 0.01 T applied perpendicular to the {\itshape c} axis, and (c) normalized electrical resistance of \CaNi samples with {\itshape T}$_{A/Q}$ = 500$^\circ$C. Susceptibility data in (a) have been offset from each other by an integer multiple of 1 $\times$ 10$^{-4}$ emu/mole for clarity.}
\label{Ni500C}
\end{figure}

Figure \ref{Ni500C} presents the data used to construct the {\itshape T}-{\itshape x} phase diagram for Ni-substitution with {\itshape T}$_{A/Q}$ = 500$^\circ$C shown in Fig. \ref{NiTxTT}a. The AFM/ORTH transition is suppressed completely between {\itshape x} = 0.017 and 0.019. Sample with {\itshape x} = 0.019 shows significant amount of diamagnetism with {\itshape T}$_{c}$ around 6~K. 

\subsection*{Ni-substitution {\itshape T}$_{A/Q}$ = 960$^\circ$C}

\begin{figure}[!htbp]
\begin{center}
\includegraphics[angle=0,width=100mm]{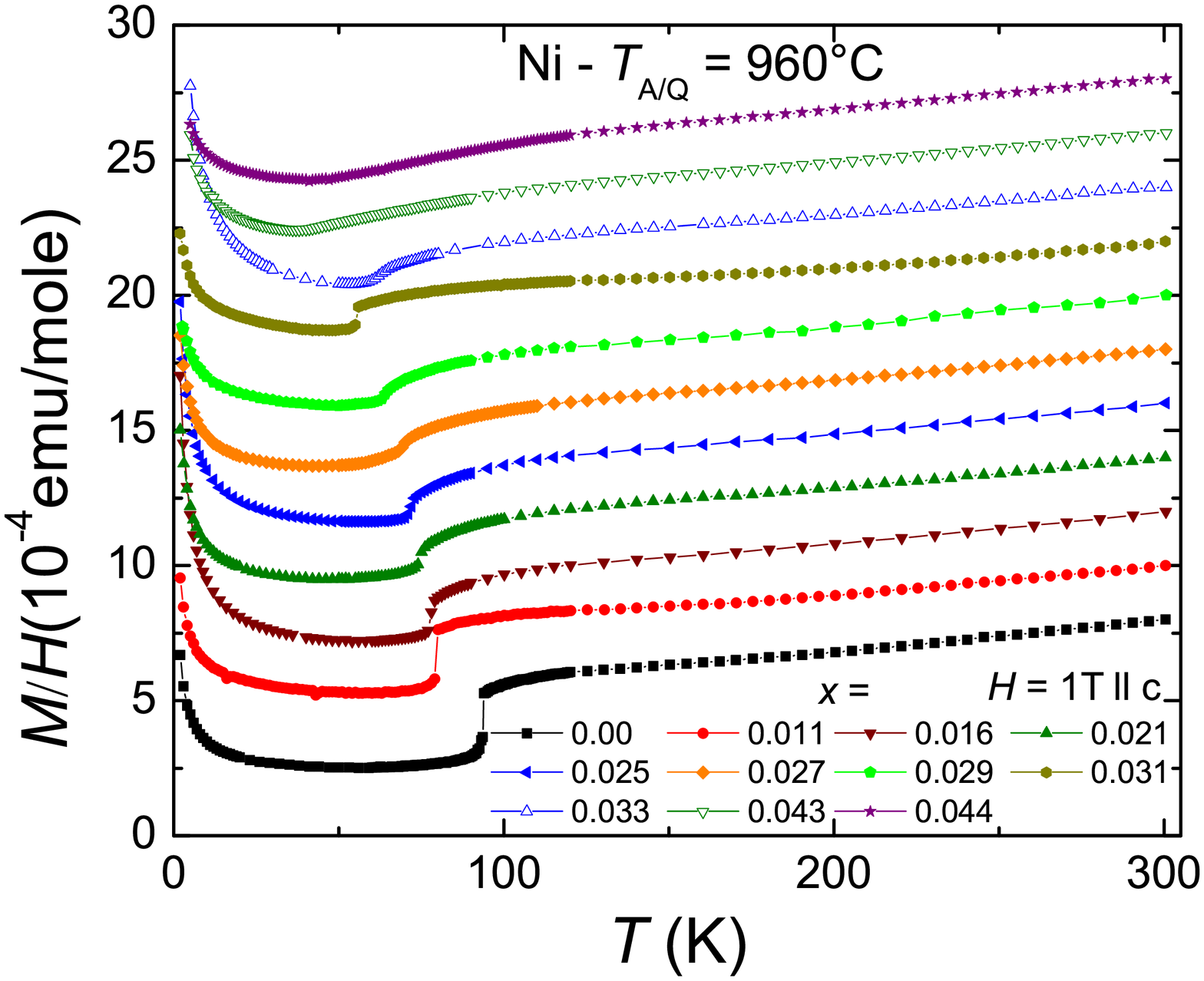}
\end{center}
\caption{(Color online) Temperature dependent magnetic susceptibility with field applied parallel to the {\itshape c} axis of \CaNi samples with {\itshape T}$_{A/Q}$ = 960$^\circ$C. Data have been offset from each other by an integer multiple of 2 $\times$ 10$^{-4}$ emu/mole for clarity. }
\label{Niasgrown}
\end{figure}

Figure \ref{Niasgrown} presents the data used to construct the {\itshape T}-{\itshape x} phase diagram for Ni-substitution with {\itshape T}$_{A/Q}$ = 960$^\circ$C shown in Fig. \ref{NiTxTT}b. The drop in susceptibility is suppressed to lower temperature as Ni substitution level is increased.

\subsection*{Ni-substitution {\itshape x} = 0.021}

\begin{figure}[!htbp]
\begin{center}
\includegraphics[angle=0,width=150mm]{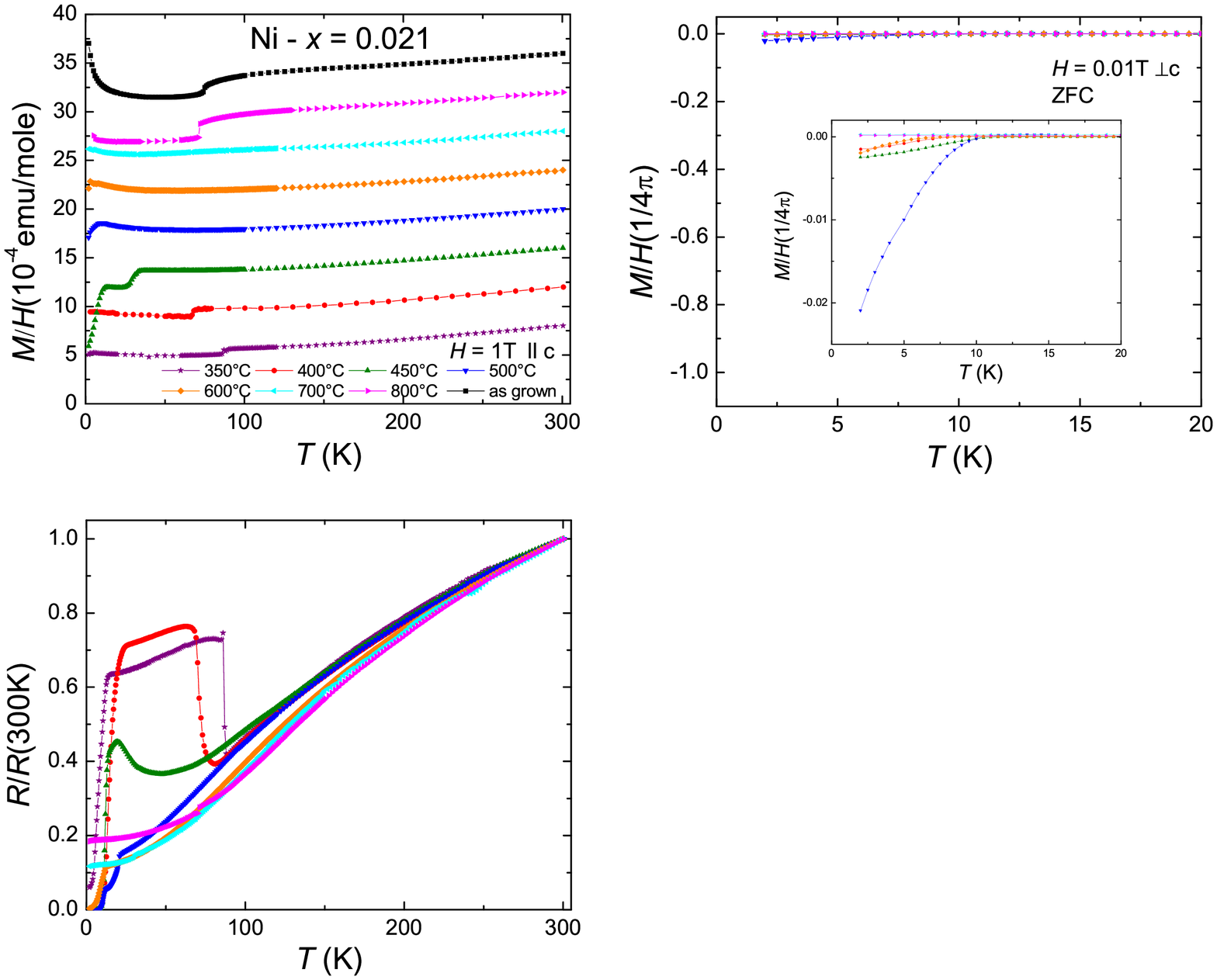}
\end{center}
\caption{(Color online) Temperature dependent (a) magnetic susceptibility with field applied parallel to the {\itshape c} axis, (b) low-field magnetic susceptibility measured upon ZFC with a field of 0.01 T applied perpendicular to the {\itshape c} axis, and (c) normalized electrical resistance of \CaNi samples with Ni concentration {\itshape x} = 0.021. Susceptibility data in (a) have been offset from each other by an integer multiple of 5 $\times$ 10$^{-4}$ emu/mole for clarity.}
\label{Ni2p1}
\end{figure}

Figure \ref{Ni2p1} presents the data used to construct the {\itshape T}-{\itshape T}$_{A/Q}$ phase diagram for Ni-substitution with {\itshape x} = 0.021 shown in Fig. \ref{NiTxTT}c. The AFM/ORTH phase transition takes place for {\itshape T}$_{A/Q}$ $\leqslant$ 450$^\circ$C and the cT phase is stabilized for {\itshape T}$_{A/Q}$ $\geqslant$ 800$^\circ$C. No bulk superconductivity is observed for any {\itshape T}$_{A/Q}$.  

\subsection*{Ni-substitution {\itshape x} = 0.026}

\begin{figure}[!htbp]
\begin{center}
\includegraphics[angle=0,width=150mm]{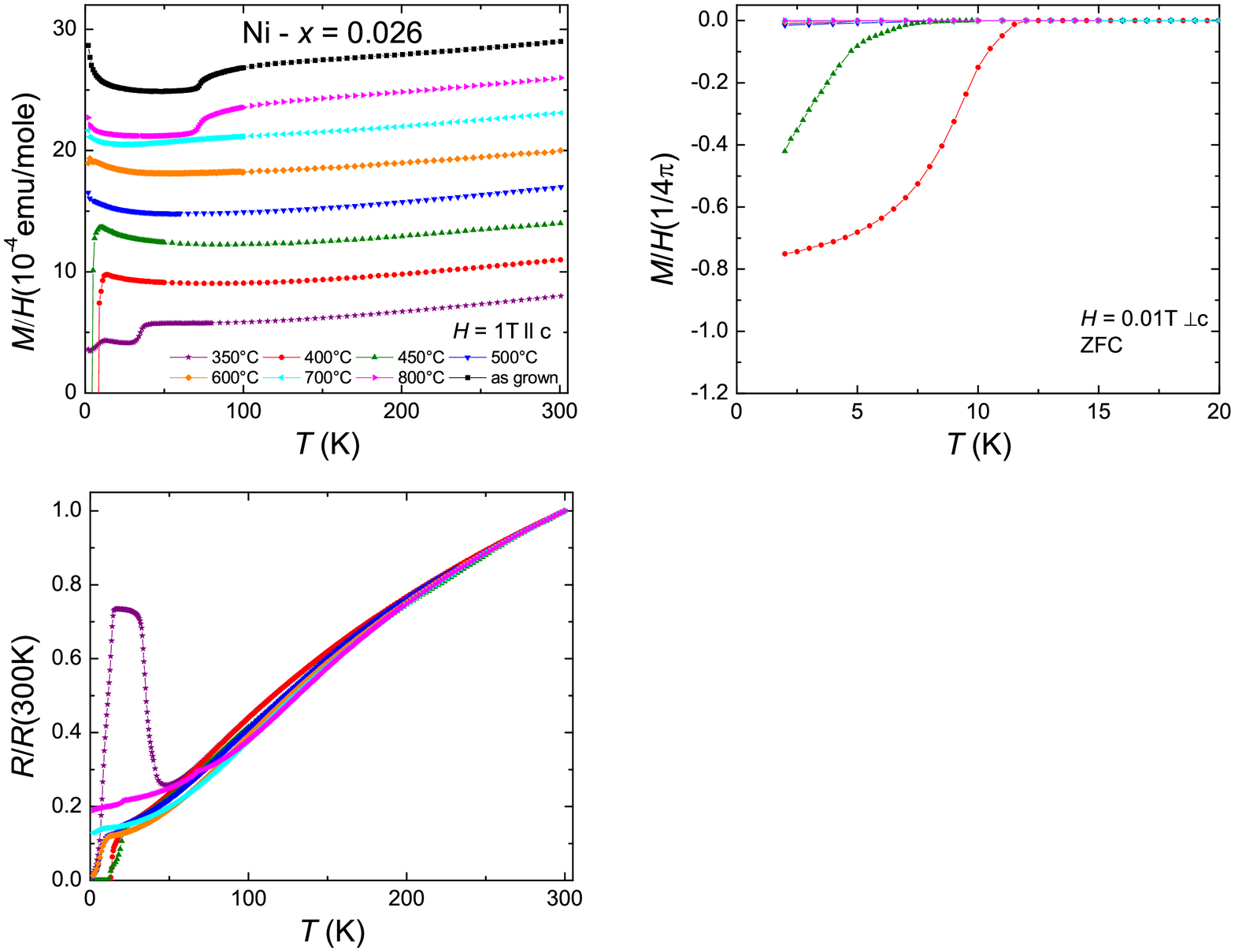}
\end{center}
\caption{(Color online) Temperature dependent (a) magnetic susceptibility with field applied parallel to the {\itshape c} axis, (b) low-field magnetic susceptibility measured upon ZFC with a field of 0.01 T applied perpendicular to the {\itshape c} axis, and (c) normalized electrical resistance of \CaNi samples with Ni concentration {\itshape x} = 0.026. Susceptibility data in (a) have been offset from each other by an integer multiple of 5 $\times$ 10$^{-4}$ emu/mole for clarity.}
\label{Ni2p6}
\end{figure}

Figure \ref{Ni2p6} presents the data used to construct the {\itshape T}-{\itshape T}$_{A/Q}$ phase diagram for Ni-substitution with {\itshape x} = 0.026 shown in Fig. \ref{NiTxTT}c. Sample with {\itshape T}$_{A/Q}$ = 400$^\circ$C shows superconductivity with screening of 80$\%$ of 1/4$\pi$. The cT phase is stabilized for {\itshape T}$_{A/Q}$ $\geqslant$ 800$^\circ$C.

\subsection*{Rh-substitution {\itshape T}$_{A/Q}$ = 500$^\circ$C}

\begin{figure}[!htbp]
\begin{center}
\includegraphics[angle=0,width=150mm]{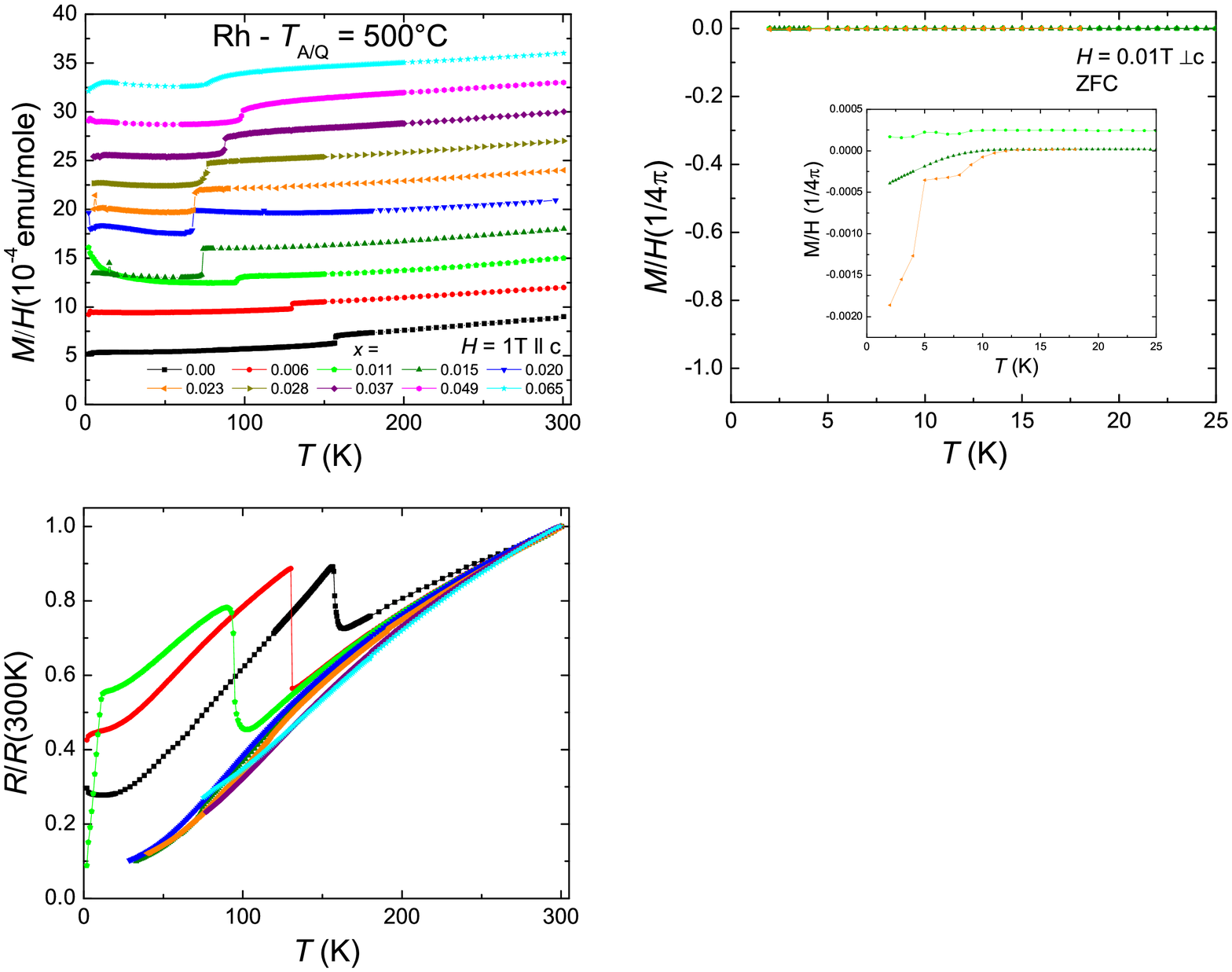}
\end{center}
\caption{(Color online) Temperature dependent (a) magnetic susceptibility with field applied parallel to the {\itshape c} axis, (b) low-field magnetic susceptibility measured upon ZFC with a field of 0.01 T applied perpendicular to the {\itshape c} axis, and (c) normalized electrical resistance of \CaRh samples with {\itshape T}$_{A/Q}$ = 500$^\circ$C. Susceptibility data in (a) have been offset from each other by an integer multiple of 3 $\times$ 10$^{-4}$ emu/mole for clarity.}
\label{Rh500C}
\end{figure}

Figure \ref{Rh500C} presents the data used to construct the {\itshape T}-{\itshape x} phase diagram for Rh-substitution with {\itshape T}$_{A/Q}$ = 500$^\circ$C shown in Fig. \ref{RhTxTT}a. The AFM/ORTH phase transition is suppressed to 95~K by {\itshape x} = 0.011 and the cT phase is stabilized by {\itshape x} = 0.015. No bulk superconductivity is observed for any Rh substitution level.

\subsection*{Rh-substitution {\itshape T}$_{A/Q}$ = 960$^\circ$C}

\begin{figure}[!htbp]
\begin{center}
\includegraphics[angle=0,width=100mm]{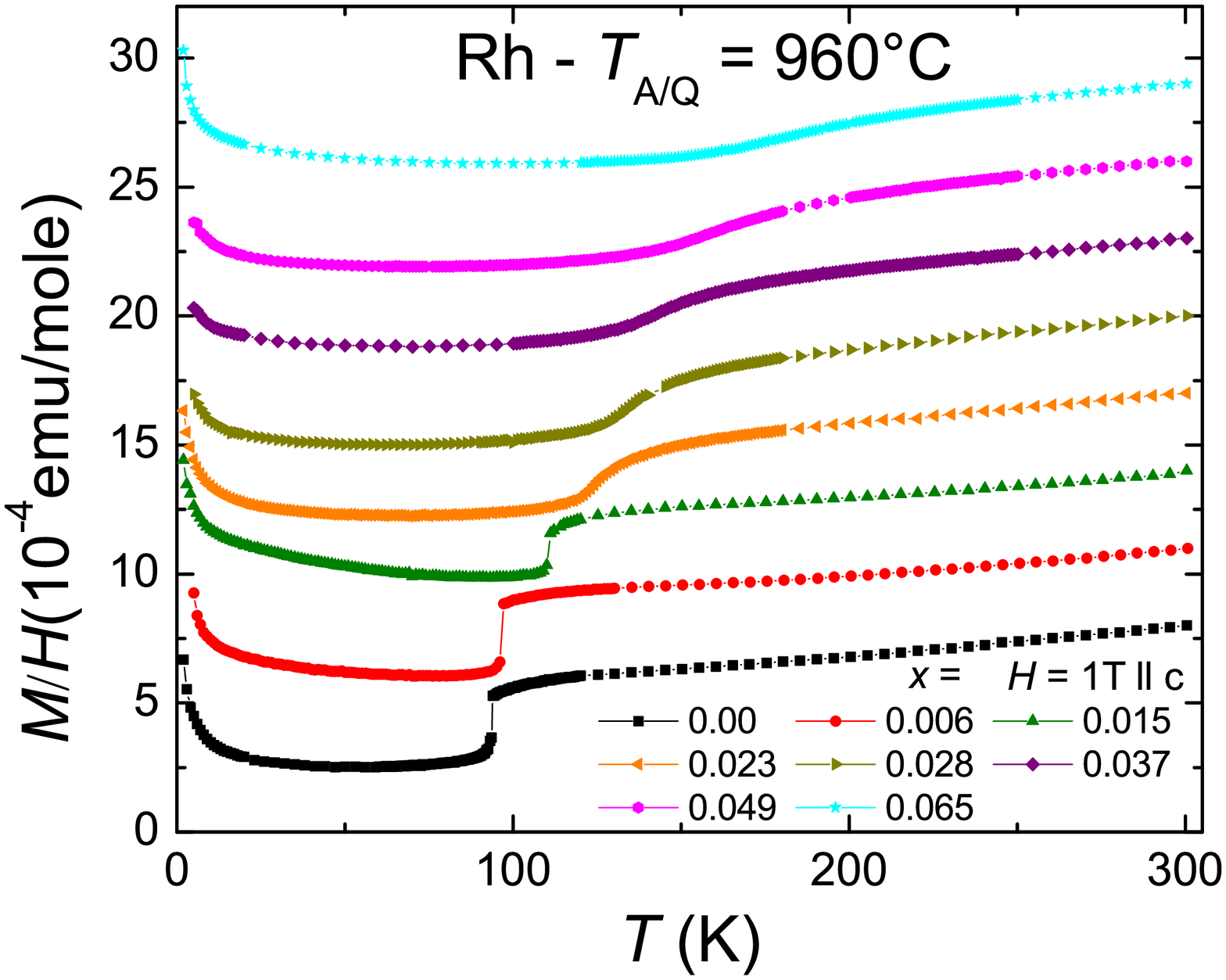}
\end{center}
\caption{(Color online) Temperature dependent magnetic susceptibility with field applied parallel to the {\itshape c} axis of \CaRh samples with {\itshape T}$_{A/Q}$ = 960$^\circ$C . Data have been offset from each other by an integer multiple of 3 $\times$ 10$^{-4}$ emu/mole for clarity.}
\label{Rhasgrown}
\end{figure}

Figure \ref{Rhasgrown} presents the data used to construct the {\itshape T}-{\itshape x} phase diagram for Rh-substitution with {\itshape T}$_{A/Q}$ = 960$^\circ$C shown in Fig. \ref{RhTxTT}b. With increasing Rh substitution level, the transition temperature of the cT phase is enhanced significantly and the feature associated with the phase transition becomes significantly broadened.

\subsection*{Rh-substitution {\itshape x} = 0.015}

\begin{figure}[!htbp]
\begin{center}
\includegraphics[angle=0,width=150mm]{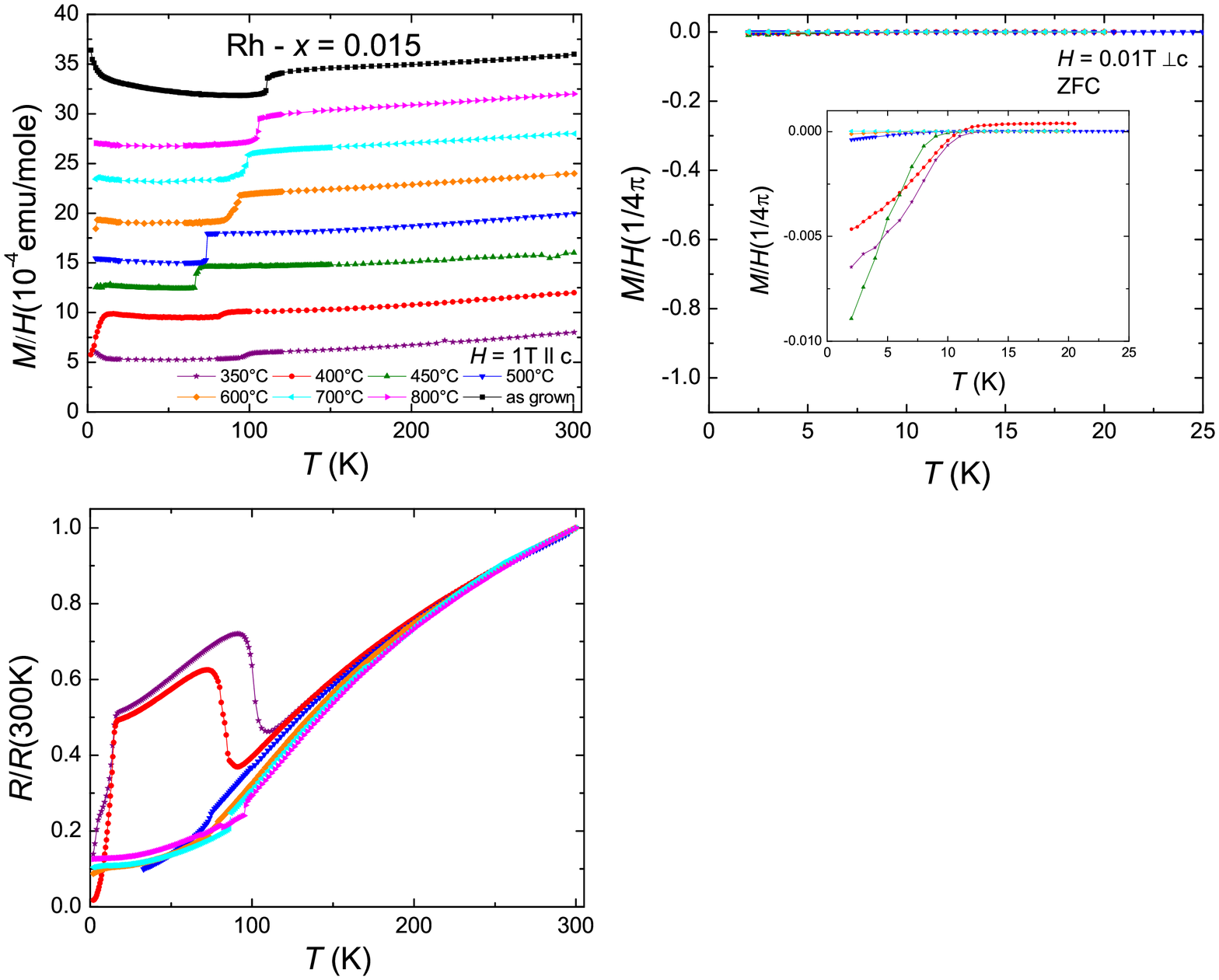}
\end{center}
\caption{(Color online) Temperature dependent (a) magnetic susceptibility with field applied parallel to the {\itshape c} axis, (b) low-field magnetic susceptibility measured upon ZFC with a field of 0.01 T applied perpendicular to the {\itshape c} axis, and (c) normalized electrical resistance of \CaNi samples with Rh concentration {\itshape x} = 0.015. Susceptibility data in (a) have been offset from each other by an integer multiple of 4 $\times$ 10$^{-4}$ emu/mole for clarity.}
\label{Rh1p5}
\end{figure}

Figure \ref{Rh1p5} presents the data used to construct the {\itshape T}-{\itshape T}$_{A/Q}$ phase diagram for Rh-substitution with {\itshape x} = 0.015 shown in Fig. \ref{RhTxTT}c. The AFM/ORTH phase transition takes place for {\itshape T}$_{A/Q}$ $\textless$ 450$^\circ$C and the cT phase is stabilized for {\itshape T}$_{A/Q}$ $\geqslant$ 450$^\circ$C. No bulk superconductivity is observed for any {\itshape T}$_{A/Q}$.

\subsection*{Rh-substitution {\itshape x} = 0.023}

\begin{figure}[!htbp]
\begin{center}
\includegraphics[angle=0,width=150mm]{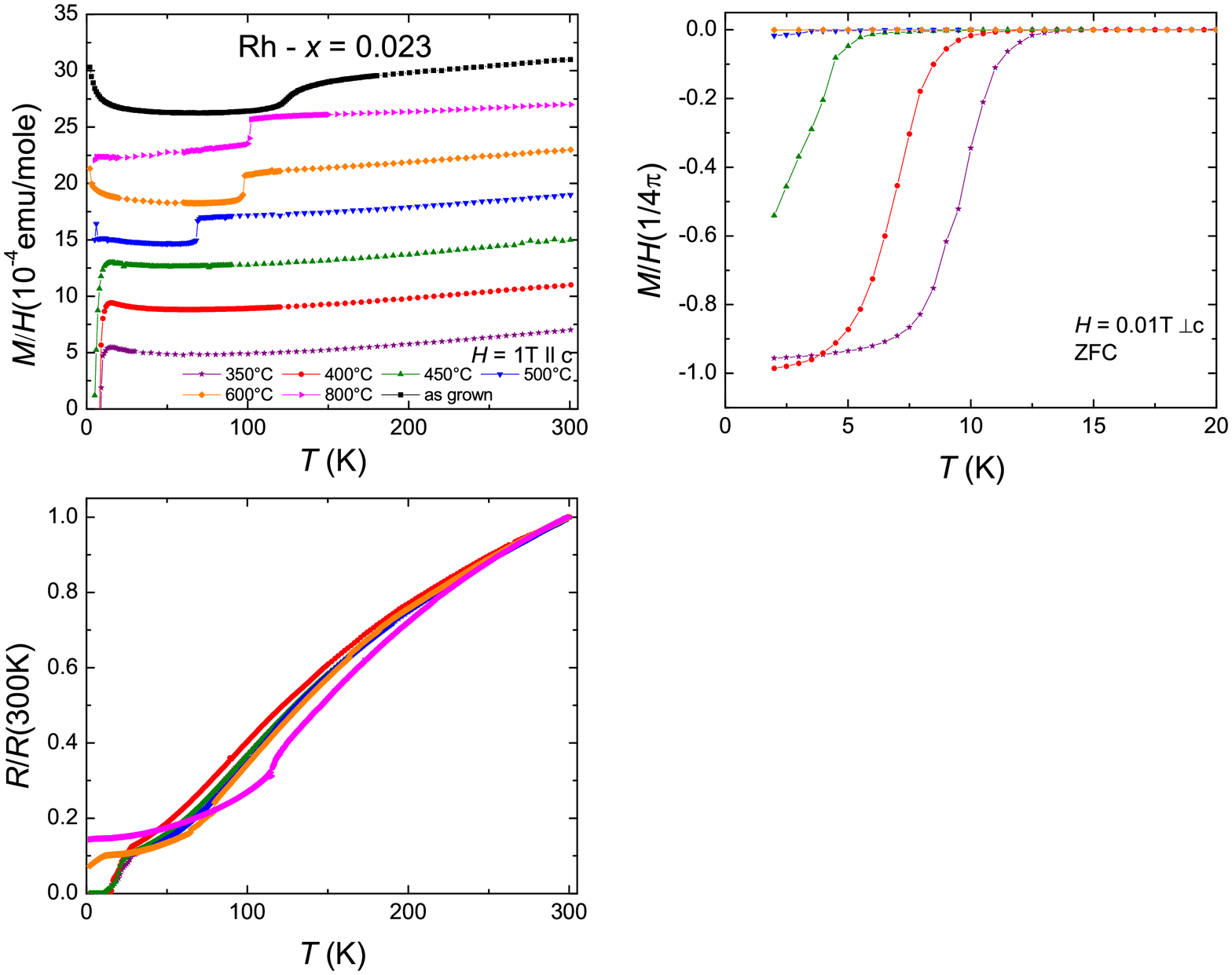}
\end{center}
\caption{(Color online) Temperature dependent (a) magnetic susceptibility with field applied parallel to the {\itshape c} axis, (b) low-field magnetic susceptibility measured upon ZFC with a field of 0.01 T applied perpendicular to the {\itshape c} axis, and (c) normalized electrical resistance of \CaNi samples with Rh concentration {\itshape x} = 0.023. Susceptibility data in (a) have been offset from each other by an integer multiple of 4 $\times$ 10$^{-4}$ emu/mole for clarity.}
\label{Rh2p3}
\end{figure}

Figure \ref{Rh2p3} presents the data used to construct the {\itshape T}-{\itshape T}$_{A/Q}$ phase diagram for Rh-substitution with {\itshape x} = 0.023 shown in Fig. \ref{RhTxTT}d. Superconductivity with full screening is observed for samples with {\itshape T}$_{A/Q}$ = 350$^\circ$C and 400$^\circ$C. The cT phase is stabilized for {\itshape T}$_{A/Q}$ $\geqslant$ 500$^\circ$C.

\label{lastpage}

\end{document}